\documentclass[fleqn,usenatbib]{mnras}

\usepackage[T1]{fontenc}
\usepackage{ae,aecompl}
\usepackage{adjustbox}

\usepackage{graphicx}   % Including figure files
\usepackage{amsmath}    % Advanced maths commands
\usepackage{amssymb}    % Extra maths symbols
\usepackage{subfigure}
\usepackage{comment}

\usepackage{soul}
\usepackage{xcolor}
\def\lsim{\mathrel{\rlap{\lower3.5pt\hbox{\hskip0.5pt$\sim$}}
    \raise0.5pt\hbox{$<$}}}

\def\gsim{~\rlap{$>$}{\lower 1.0ex\hbox{$\sim$}}}

\def\mst{\mbox{$M_{\star}$}}

\definecolor{gda}{rgb}{1,1,1}
\title[LAMOST-DR7 velocity dispersion catalog]{Central velocity dispersion catalog of LAMOST-DR7 galaxies}

\author[N.R. Napolitano et al.]{
Nicola R. Napolitano$^{1}$\thanks{E-mail: napolitano@mail.sysu.edu.cn (NRN)},
Giuseppe D'Ago$^{2,3}$,
Crescenzo Tortora$^{4}$,
Gang Zhao$^{5,6}$,
\and
A-Li Luo$^{5,6}$,
Baitian Tang$^{1}$,
Wei Zhang$^{5}$,
Yong Zhang$^{7}$,
Rui Li$^{1}$\\
\\
$^{1}$School of Physics and Astronomy, Sun Yat-sen University, Zhuhai Campus, 2 Daxue Road, Xiangzhou District, Zhuhai 519082, China\\
$^{2}$Instituto de Astrofísica, Pontificia Universidad Católica de Chile, Av. Vicuña Mackenna 4860, 7820436 Macul, Santiago, Chile\\
$^{3}$INAF-Osservatorio Astronomico di Capodimonte, Salita Moiariello 16, I-80131 Napoli\\
$^{4}$INAF -- Osservatorio Astrofisico di Arcetri, Largo Enrico Fermi 5, 50125, Firenze, Italy\\
$^{5}$CAS Key Laboratory of Optical Astronomy, National Astronomical Observatories, Chinese Academy of Science, Beijing 100101, China\\
$^{6}$School of Astronomy and Space Science, University of Chinese Academy of Sciences,  Beijing 100049, China\\
$^{7}$Nanjing Institute of Astronomical Optics \& Technology, National Astronomical Observatories, Chinese Academy of Sciences, Nanjing \\210042, China
}

\date{Accepted XXX. Received YYY; in original form ZZZ}

\pubyear{2020}

\begin{document}
\label{firstpage}
\pagerange{\pageref{firstpage}--\pageref{lastpage}}
\maketitle

\begin{abstract}
The Large Sky Area Multi-Object Fiber Spectroscopic Telescope (LAMOST) is a major facility to carry out spectroscopic surveys for cosmology and galaxy evolution studies. The seventh data release of the LAMOST ExtraGAlactic Survey (LEGAS) is currently available and including redshifts of 193\,361 galaxies. These sources are spread over $\sim 11\,500$ deg$^2$ of the sky, largely overlapping with other imaging (SDSS, HSC) and spectroscopic (BOSS) surveys. The estimated depth of the galaxy sample, $r\sim17.8$, the high signal-to-noise ratio, and the spectral resolution $R=1800$, make the LAMOST spectra suitable for galaxy velocity dispersion measurements, which are invaluable to study the structure and formation of galaxies and to determine their central dark matter (DM) content.
We present the first estimates of central velocity dispersion of $\sim86\,000$ galaxies in LAMOST footprint. We have used a wrap-up procedure to perform the spectral fitting using \textsc{pPXF}, and derive velocity dispersion measurements. Statistical errors are also assessed by comparing LAMOST velocity dispersion estimates with the ones of SDSS and BOSS over a common sample of $\sim51\,000$ galaxies. The two datasets show a good agreement, within the statistical errors, in particular when velocity dispersion values are corrected to 1 effective radius aperture. We also present a preliminary Mass-$\sigma$ relation and find consistency with previous analyses based on local galaxy samples.
These first results suggest that LAMOST spectra are suitable for galaxy velocity dispersion measurements to complement the available catalogs of galaxy internal kinematics in the northern hemisphere. We plan to expand this analysis to next LAMOST data releases.
\end{abstract}


%%%%%%%%%%%%%%%%%%%%%%%%%%%%%%%%%%%%%%%%%%%%%%%%%%

%%%%%%%%%%%%%%%%% BODY OF PAPER %%%%%%%%%%%%%%%%%%
\section{INTRODUCTION}
\label{sec:introduction}

Spectroscopic surveys are fundamental tools to determine galaxy distances through their redshifts and map their distribution across time and space, e.g. to reconstruct their large scale structure (see e.g., {the 2dF Galaxy Redshift Survey}, 2dFGRS, \citealt{2df2001}; {the 6dF Galaxy Redshift Survey}, 6dFGRS, \citealt{6df2009}; {the Baryon Oscillation Spectroscopic Survey}, BOSS, \citealt{boss2013}; {the VIMOS Public Extragalactic Redshift Survey}, VIPERS, \citealt{vipers2017}) and investigate their assembly (e.g. {the Galaxy Mass Assembly}, GAMA, \citealt{2015MNRAS.452.2087L}) and the evolution of their properties and clustering up to high redshift (e.g. the DEEP2 survey, \citealt{2013ApJS..208....5N}). Additionally, they are also used to calibrate the tools to derive photometric redshifts \citep[see e.g., ][]{cavuoti2015,cavuoti2017}, which are invaluable for imaging surveys devoted to cosmological studies, making use, e.g., of weak gravitational lensing (e.g., \citealt{hildebrandt2017}), baryonic acoustic oscillation (e.g., \citealt{Ahn+12_SDSS_DR9}; \citealt{Bautista+18_BAO}) or large scale structure (e.g., \citealt{Alpaslan2015}).

%lee2018

Spectroscopic redshifts are also crucial for galaxy formation studies as they allow to accurately select galaxies at different cosmic epochs to study the main scaling relations such as size-luminosity  and size-mass \citep{shen2003,HB09_curv}, Kormendy relation \citep[KR, ][]{Kormendy1977,HamabeKormendy1987}, and trace these relations back in time \citep[see e.g.,][]{huertas2013,Roy+18}.\\

%The visible wavelength range is largely investigated for dark matter studies through gravitational lensing \citep[e.g., ][]{hildebrandt2017,spiniello2018}, for the large scale structure formation \citep[][]{Alpaslan2015} and galaxy evolution studies \citep{bernardi2009,lee2018,roy2018} and the upcoming Large Synoptic Survey Telescope \citep[LSST, ][]{lsst2008} will revolutionise the knowledge in these fields from the 2020s and in the decades ahead.\\
%All the aforementioned studies are based on photometry. While photometric surveys are the best and most efficient way to study the global properties of the Universe by imaging large portions of the sky, it is crucial to perform spectroscopic follow-ups, in order to better investigate the volume and the three-dimensional distribution of galaxies.

Besides galaxy redshifts, spectroscopic surveys can also provide galaxy kinematics (see e.g. \citealt{2013MNRAS.431.1383T}), hence allowing to investigate more fundamental scaling relations like the Faber-Jackson and the mass--dispersion relation (\citealt{HB09_curv}), and the fundamental plane (e.g., \citealt{SPIDER-II}), investigating the correlation between central DM fraction, mass density slope, Initial Mass Function (IMF) and mass (\citealt{Cappellari+06,Cappellari+13_ATLAS3D_XX}; \citealt{tortora2009,SPIDER-VI,TRN13_SPIDER_IMF,Tortora+14_DMslope,Tortora+14_DMevol, Tortora+18_KiDS_DMevol,NRT10}; \citealt{CT10}; \citealt{ThomasJ+09,ThomasJ+11}; \citealt{Dutton_Treu14}) and trace them back in time \citep[see e.g.,][]{Tortora+10lensing,Tortora+18_KiDS_DMevol,Beifiori+14} or probe alternative gravities (\citealt{Tortora+14_MOND,Tortora+18_Verlinde}).

%or  investigating the dark matter fraction in galaxy centers (xxxx) as a function of redshift (xxxx).

\begin{figure*}
    \centering
        \includegraphics[width=2\columnwidth]{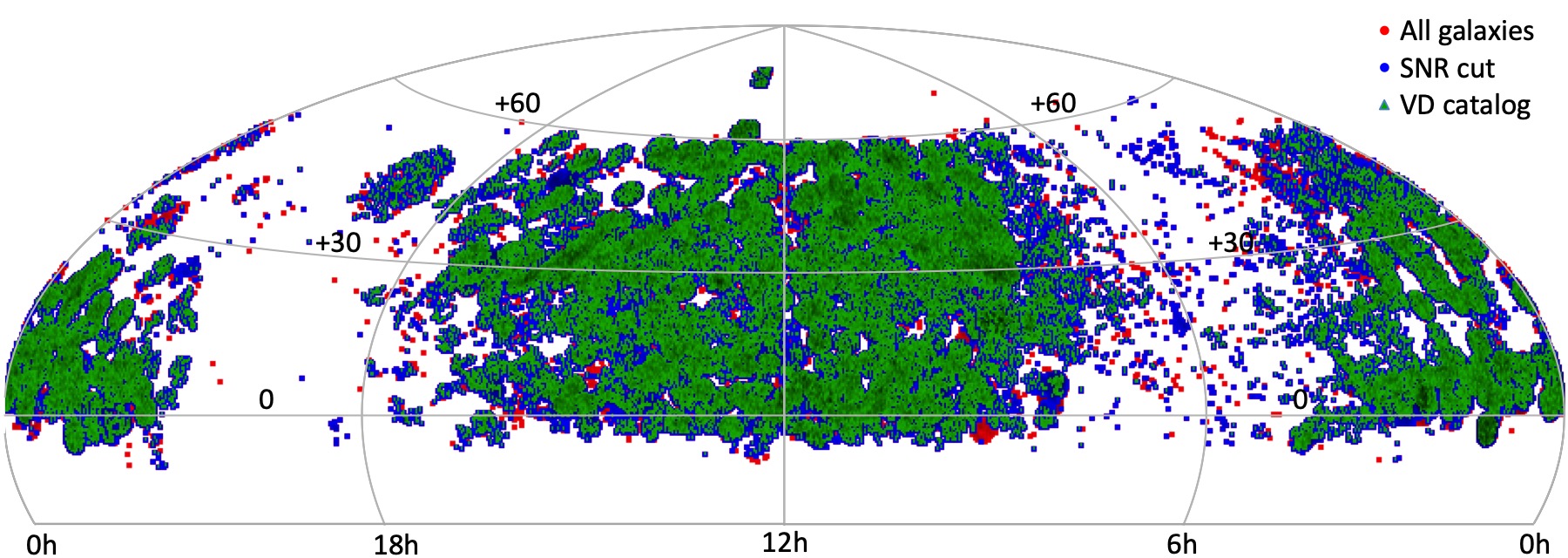}
    \label{fig:LAMOST_sky}
    \caption{Sky distribution (RA, DEC) of LAMOST galaxy targets from the LEGAS survey. In red all sources labeled as galaxies, in blue the ones that have passed the SNR criteria (i.e. $SNR>10$ in $g$- or $i$-band) and in green the galaxies with measured velocity dispersion in this work.}
\end{figure*}

These studies are especially valuable if one can combine internal kinematics with high image quality data, to determine accurate light profile of galaxies to be used to derive their central dynamics (see e.g. \citealt{Beifiori+14}; \citealt{Tortora+18_KiDS_DMevol}), to study the galaxy-black hole co-evolution (e.g. \citealt{SDSS_BH_sigma}), and to combine the gravitational lensing with the internal kinematics of the lens to obtain constraints on the dark matter properties (e.g., \citealt{Cardone+09}; \citealt{CT10}; \citealt{Auger+10_SLACSX}; \citealt{Tortora+10lensing}) and the galaxy Initial Mass Function (e.g. \citealt{Auger+10}; \citealt{Sonnenfeld+15_SL2SV}; \citealt{2015MNRAS.452L..21S}).

The accumulation of larger area of the sky with such a high-quality imaging data, both in Northern sky, e.g. with the {Hyper Suprime-Cam Subaru Strategic Program (HSC,  \citealt{2012SPIE.8446E..0ZM}; \citealt{Aihara+18_HSC}), and in the Southern hemisphere, e.g. with the Kilo Degree Survey, KiDS and its twin near infrared VIKING survey (see e.g. \citealt{Kuijken+19_KiDS-DR4}), or the Dark Energy Survey \citep[DES,][]{2005astro.ph.10346T},} is making particularly useful to catch up with equally extended and possibly deep spectroscopic datasets from which galaxy kinematics can be extracted. For this reason, future surveys are specifically designed to cover this gap (e.g., WAVES@4MOST, \citealt{Driver+16_WAVES}, STEPS/WEAVE@WHT \citealt{2019A&A...632A...9C}).

%Addittionally, they are also used to calibrate the tools to derive photometric redshifts \citep[see e.g., ][]{cavuoti2015,cavuoti2017}. To be highly valuable for galaxy formation studies, spectroscopic surveys have to reach the highest completeness possible in order to allow to obtain unbiased constraints on the main scaling relations such as size-luminosity \citep{Hyde2009}, size-mass \citep{shen2003,Hyde2009}, Kormendy relation \citep[KR, ][]{Kormendy1977,HamabeKormendy1987} and trace these relations back in time \citep[see e.g., ][]{huertas2013,roy2018}.\\
%NRN alcune di queste xxx le trovi in Roy et al., ti suggerisco di leggere quel paper

The Large sky Area Multi-Object Fiber Spectroscopic Telescope (LAMOST) is a 4m effective aperture (6.67m main mirror sized) telescope located at the Xinglong Observatory northeast of Beijing, China (\citealt{1996ApOpt..35.5155W,Cui+12_LAMOST,Zhao+12_LAMOST}). The telescope is characterized by a $\sim 20 $deg$^2$ field of view (FOV) where a total of 4000 circular fibers of 3.3 arcsec diameter can be mounted on its focal plane (\citealt{Deng+12_LAMOST-LEGUE}). With the capability of its large field of
view and strong multiplexing ability, the LAMOST telescope is dedicated to a spectral survey of celestial objects over the entire available northern sky. LAMOST is a Chinese national scientific research facility operated by National Astronomical Observatories, Chinese Academy of Sciences (NAOC). Since the commissioning year started in 2009, LAMOST has carried out a spectral survey of millions of objects in a large part of the northern sky. The survey has two major components: the LAMOST ExtraGAlactic Survey (LEGAS) and the LAMOST Experiment for Galactic Understanding and Exploration (LEGUE).

Currently the survey has collected observations for one year pilot survey followed by eight year regular survey, and the seventh data release (DR7 hereafter) has been domestically delivered including the data of the first 7 years of regular survey plus the pilot survey.

In particular, the extragalactic component (LEGAS), has collected spectra for 193\,361 galaxies  and 64\,236 quasars
(see also \S\ref{sec:LEGAS}). Like previous data releases, for all these sources data LAMOST-DR7 provides flux- and wavelength-calibrated, sky-subtracted spectra and a series of basic spectro-photometric information like right ascension, declination, signal-to-noise ratio (SNR hereafter), magnitude, source classification and redshift (see e.g. \citealt{2015RAA....15.1095L}).

These data have been already used for extragalactic science like the study of E+A galaxies (\citealt{2015RAA....15.1414Y}), the analysis of quasar properties (\citealt{2016AJ....151...24A, 2014A&A...564A..89S}), the discovery and analysis of double peaked AGNs (\citealt{2014MNRAS.439.2927H,2014RAA....14.1234S}, \citealt{2019MNRAS.482.1889W}), the study of Luminous Infrared Galaxies (\citealt{2015RAA....15.1424L}), the detection and analysis of galaxy pairs (\citealt{2016RAA....16...43S}, \citealt{2019ApJ...880..114F}), the study of compact groups (\citealt{2020ApJS..246...12Z}), the spectral classification and composites of galaxies (\citealt{2018MNRAS.474.1873W}). Within the extragalactic program there is also the LAMOST Complete Spectroscopic Survey of Pointing Area (LaCoSSPAr) in the Southern Galactic Cap (\citealt{2018ApJS..234....5Y}, \citealt{2019RAA....19..113Z}), which is the only area (two fields of 20 deg$^2$ each) with a magnitude limited sample ($r=18.1$\footnote{All magnitudes reported in this paper are in AB system.}).

Deriving the velocity dispersion for a large subsample of LAMOST galaxies would represent a big scientific value, for instance: 1) to correlate the galaxy properties (e.g. luminosity, mass-to-light ratios, sizes, star formation rate) with a proxy of the halo mass, the velocity dispersion (e.g., \citealt{HB09_FP,HB09_curv}), 2) to estimate DM fraction, mass density slope and IMF and their evolution with redshift \citep[see e.g., ][]{Cappellari+06,Cappellari+13_ATLAS3D_XX,tortora2009,TRN13_SPIDER_IMF, Tortora+14_DMslope,Tortora+18_KiDS_DMevol,Beifiori+14}; 3) to provide further and independent constraints on the total mass within the Einstein radius in strong-lensed systems (\citealt{CT10,Tortora+10lensing}; \citealt{Sonnenfeld+15_SL2SV}), 4) to test DM models or alternative gravities (e.g., \citealt{Tortora+14_MOND,Tortora+18_Verlinde}) 5) to discover compact and massive galaxies, which are expected to have large velocity dispersions (\citealt{Saulder+15_compacts}), avoiding the systematics in the galaxy size measurement, induced by the low spatial resolution of wide-field ground-based surveys (e.g., \citealt{Tortora+18_UCMGs}).

In this work, we present the first catalog of galaxy central velocity dispersion (VD hereafter) measurements from spectra of galaxies in the LAMOST-DR7. For all calculations, we assume a cosmology with ($\Omega_M$, $\Omega_\Lambda$, $h$)=(0.3, 0.7, 0.7).

%NRN below is to be completed
In \S\ref{sec:spectroscopic_database} we briefly describe the LAMOST spectroscopic database used and the selections made to perform the VD measurements; in \S\ref{sec:veldisp_measurements} we briefly illustrate the wrap-up procedure to perform the spectral fitting to derive the galaxy internal kinematics and introduce the the first LAMOST velocity dispersion catalog; in \S\ref{sec:results}  we compare the LAMOST velocity dispersion estimates with the {Sload Digital Sky Survey (SDSS)} and BOSS and discuss the relative scatter. In \S\ref{sec:mass-sigma} we introduce a preliminary analysis of the Mass-$\sigma$ relation using LAMOST galaxies for which we have found mass estimates from litarature. Finally, we discuss some perspective on the application of our procedure to next LAMOST data releases and draw some conclusions on \S\ref{sec:conclusions}

\section{The spectroscopic data}
\label{sec:spectroscopic_database}
\subsection{The LAMOST ExtraGAlactic Survey}
\label{sec:LEGAS}
The LAMOST ExtraGAlactic Survey (LEGAS) is the key project in the LAMOST extragalactic mission. It is divided into two large patches in the northern galactic cap (NGC) and the southern galactic cap (SGC) and it targets pre-selected galaxies and QSOs. The NGC patch covers about 8000 deg$^2$ and has the same footprint as the SDSS Legacy Survey. It is aimed at observing SDSS galaxies (with $r\leq$ 17.8) that
were not spectroscopic followed-up to avoid fiber collisions (\citealt{Stoughton+02_SDSS}). The SGC patch covers about 3500 deg$^2$. In this latter region ($b$ < -30$^{\circ}$ and $\delta$ > -10$^{\circ}$ in a strip of equatorial coordinates 45$^{\circ} <$  RA$ < 60^{\circ}$ and 0.5$^{\circ}<$ DEC $< 9.5^{\circ}$) LAMOST aims at taking spectra of galaxies with limiting magnitude $r < 18$, and a sample of blue galaxies down to $r < 18.8$.
%NRN: checked with KIDS MAG_GAAP and limiting magnitude might be 0.2 fainter for blue galaxies, but quite ok for all galaxies. Not sure about completeness magnitude, which might be 0.5 fainter
Besides these two
well-defined samples of galaxies, some bright-infrared galaxies are chosen as extra observational targets that are selected from infrared surveys such as IRAS, WISE, and Herschel.

So far, the LAMOST consortium has delivered 7 public data releases, the last one, DR7, in March 2020.
LAMOST-DR7 includes all targets delivered in previous releases, thus we will analyse 193\,361 spectra classified as galaxies in the LAMOST catalog.
We remark that this spectra database contains a number of re-visits, hence the actual number of observed galaxies is 163\,765. We will retain the repeated spectra to evaluate the internal consistency of the fitting procedure and/or pick the highest quality among the repeated measurements (see \ref{sec:repeated}).

\begin{figure}
%\hspace{-1cm}
        \includegraphics[width=1\columnwidth]{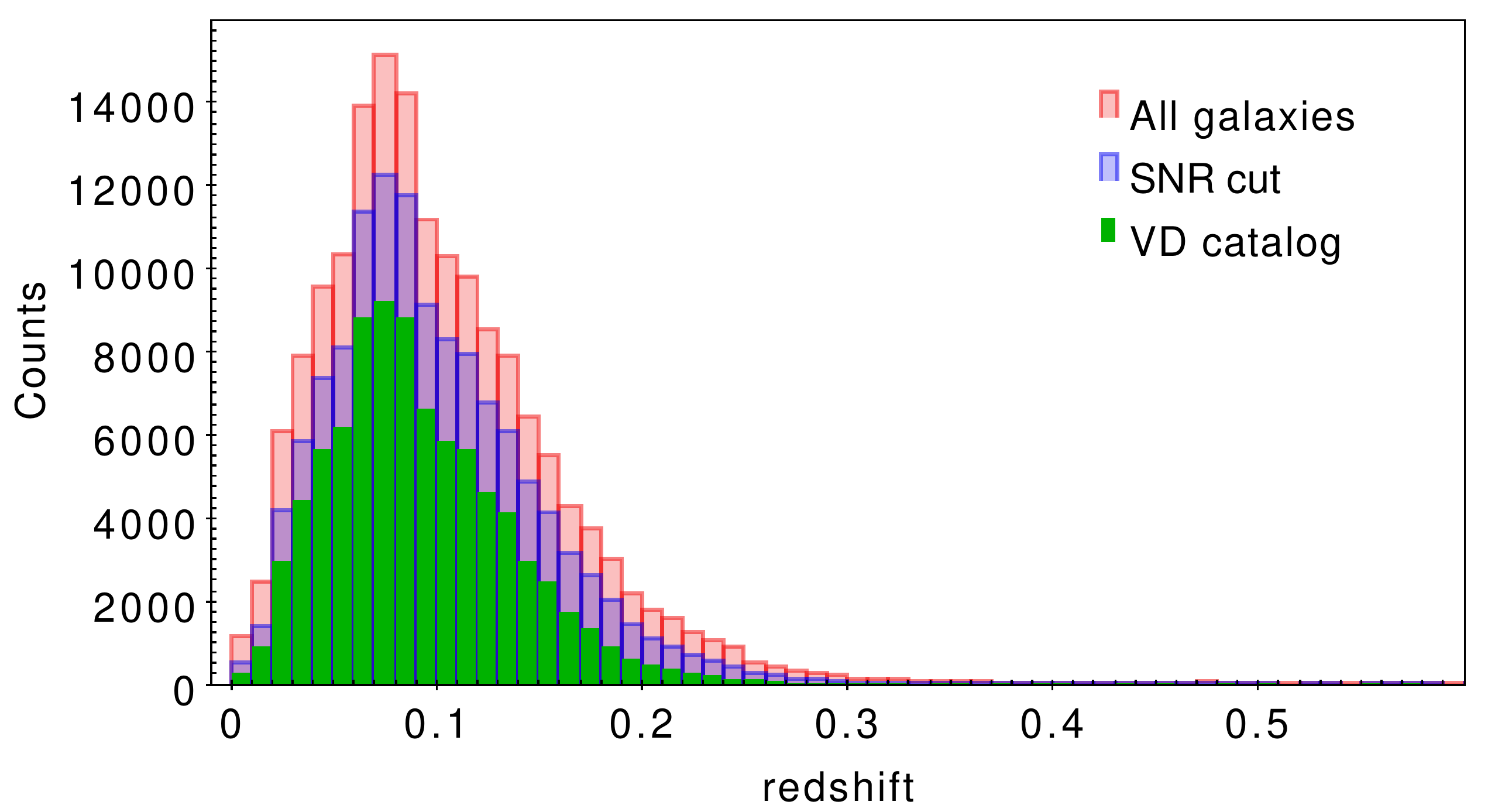}
%\hspace{-1cm}
        \includegraphics[width=1\columnwidth]{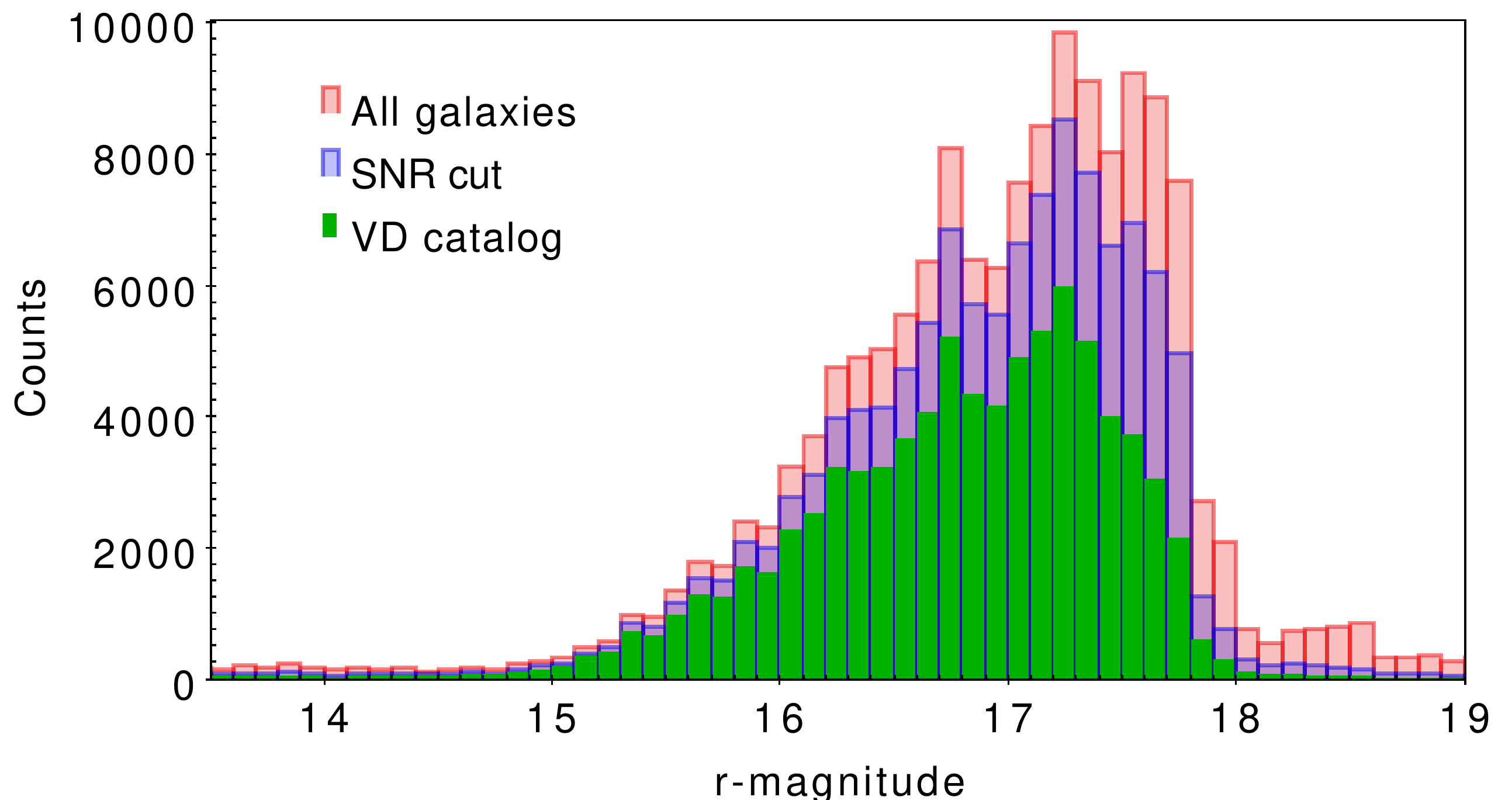}

    \caption{Redshift (top) and $r-$band magnitude (bottom) distributions of LAMOST galaxies from the LEGAS survey. In red all sources labeled as galaxies, in blue the ones that have passed the SNR criteria (i.e. $SNR>10$ in $g$- or $i$-band) and in green the galaxies with measured velocity dispersion in this work. Note that both redshift and magnitude distributions have long tails that have been cut in the figures for graphical reasons. %distributions for both the whole and the selected GAMA/AAT samples are shown in the histogram above. The GAMA/AAT galaxy sample has a relatively low SNR, for which the median value is 4.4, and we see that the cut performed on the whole GAMA/AAT sample does not change the shape of the distribution, meaning that we are not introducing selection biases in our sample.
    \label{fig:LAMOST_z}
    }
\end{figure}

The sky distribution of the
%the extragalactic sources selected as
LAMOST-DR7 galaxies
%and quasars are
is shown in Fig. \ref{fig:LAMOST_sky}. Here we plot: 1) all galaxies currently available in the LAMOST database (in red), 2) the ones with high signal-to-noise ratio (in blue, see \S\ref{sec:spectra}), and 3) the galaxies for which we have obtained VD measurements (in green, see \S\ref{sec:veldisp_measurements}). {As mentioned in \S\ref{sec:introduction}, the LaMOST-DR7 provides basic information for each galaxy in the catalog, including position, magnitude in 5 optical bands ($ugriz$), spectral SNR in the same bands, and the redshift with error derived by the LAMOST spectra (see \citealt{2015RAA....15.1095L}). In Fig. \ref{fig:LAMOST_z} we show the distribution of redshift, $z$\, and $r-$band magnitude, $mag_r$, for the three galaxy subsets introduced above.}
%The distribution of redshift ($z$) and $r-$band magnitude ($mag_r$), as provided by LAMOST catalogs, are shown in Fig. \ref{fig:LAMOST_z} for the same galaxy subsets.
In particular, LAMOST galaxies are mainly concentrated in the redshift range $z=$[0,0.3] and they have a mean redshift $\langle z \rangle=0.10\pm0.06$ (where the error is the  1$\sigma$ scatter of the sample distribution). The $r-$band magnitude distribution mainly covers the range $mag_r=$[14.0,19.0] with a mean $\langle mag_r \rangle=17.0\pm0.8$, and starts
%that the galaxy sample start
to become highly incomplete for $mag_r>17.8$, {due to the adopted selection function (see above)}, although the higher SNR sample show a slightly brighter limiting magnitude (see \ref{sec:spectra}).
%In the same figure we also show the footprint of the major overlapping optical imaging
%SDSS, HSC, KiDS, DES) and spectroscopic (BOSS, GAMA) surveys.

\subsection{LAMOST spectra}
\label{sec:spectra}
The LAMOST designed wavelength coverage is 3650-9000 \AA\ and the resolution is $R\sim1800$ increased from $R\sim1000$ through narrowing the slit at the output end of the fibers to 2/3 the fiber for all 16 LAMOST spectrographs (see \citealt{2015RAA....15.1095L}).

On the original CCD image, 4K pixels are used for recording blue or red wavelength regions ranging from 3700 \AA\ to 5900 \AA\ or from 5700 \AA\ to 9000 \AA\ respectively. When the blue and red channels are combined together, each spectrum is re-binned to a resolution element of 69~km~s$^{-1}$, which means the difference in wavelength between two adjacent points
is $\log (\lambda) = 0.0001$. Conventionally this resolution element is referred to as a ``pixel'' for LAMOST data (see also \citealt{2015RAA....15.1095L} for further details).

\begin{figure}
    \centering
        \includegraphics[width=\columnwidth]{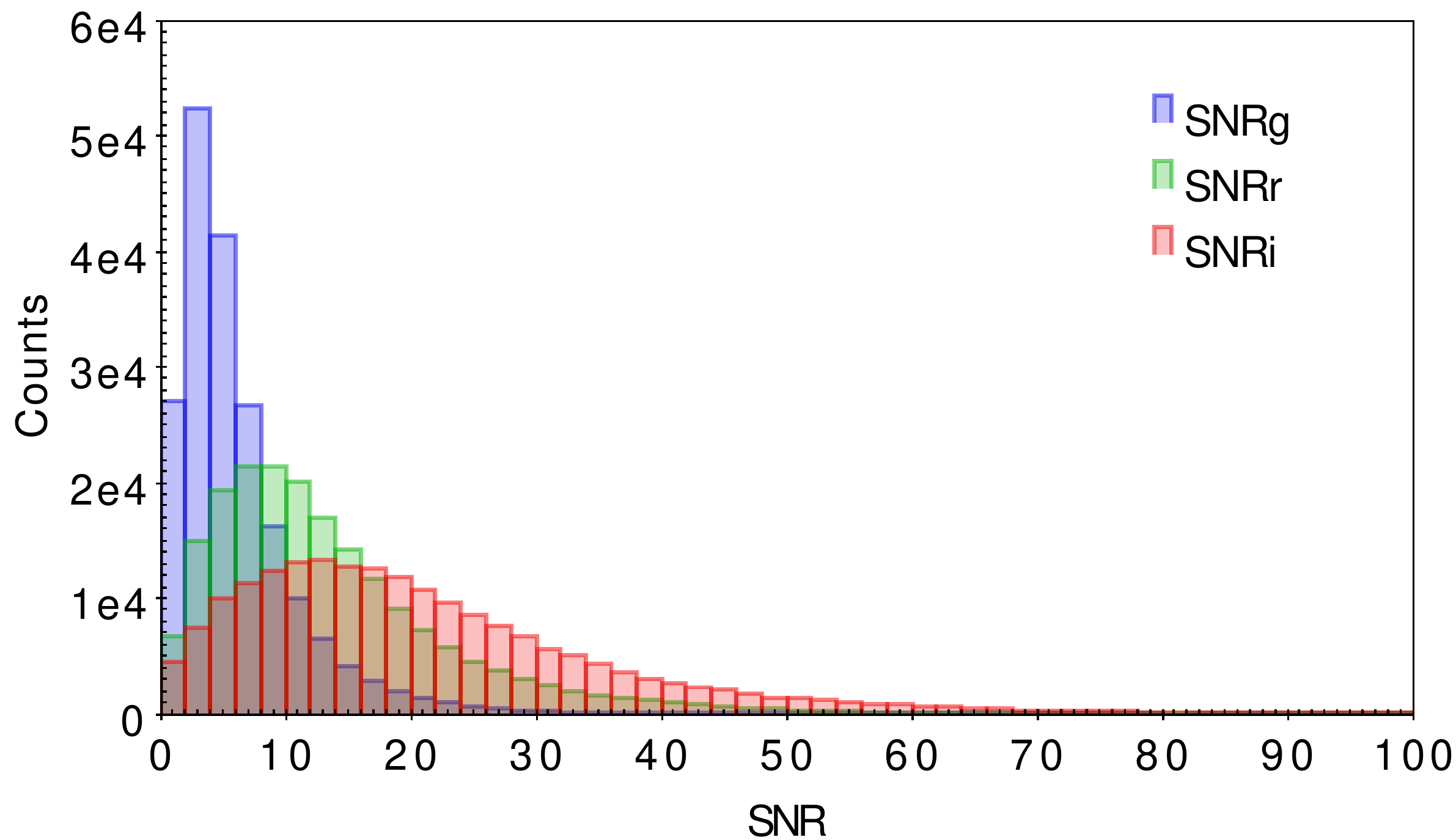}
    \caption{Signal to noise ratio distribution. The SNR/pixel as from the LAMOST galaxy catalog in $g-$band (blue), $r-$band (green) and $i-$band (red).  %distributions for both the whole and the selected GAMA/AAT samples are shown in the histogram above. The GAMA/AAT galaxy sample has a relatively low SNR, for which the median value is 4.4, and we see that the cut performed on the whole GAMA/AAT sample does not change the shape of the distribution, meaning that we are not introducing selection biases in our sample.
    }
\label{fig:LAMOST_sn_distribution}
\end{figure}

The SNR is defined per `pixel' and calculated by using the inverse-variance, which is included in the
spectral FITS files, in different band passes ($u,~g,~r,~i,~z$). The SNR for each pixel is given by the product between
the flux and its inverse-variance (\citealt{2015RAA....15.1095L}). The mean SNR in a wavelength band can be averaged from the SNR of each pixel in the bandpass range. Another way to obtain SNR is to calculate the ratio between the continuum and the 1$\sigma$ (standard deviation) of the continuum noise. These two definitions of SNR are given in the LAMOST catalogs to provide a quality flag for LAMOST data (see Fig. \ref{fig:LAMOST_sn_distribution}){, while they do not reflect the local SNR of the spectral features that are used to perform the velocity dispersion measurements. This latter SNR will be determined by the fitting procedure and used as a further quality criterion (see \S\ref{sec:veldisp_measurements}). According to the data release documentation (see \citealt{2015RAA....15.1095L})
%The data quality is determined by the SNR, and
a successful observational target with optimal data quality is defined if the ``global'' SNR$>10$ in $g$- or $i$-band. In the following we will adopt the same definition, although we plan to investigate less stringent constraints in future analyses}. {As it can be seen in Fig. \ref{fig:LAMOST_sn_distribution}, this condition is realised in $i-$band for the majority of spectra, while there is a shorter tail of galaxies with $g$-band, so the $i-$band SNR will be one mostly qualifying spectra for VD measurements.}
%The distribution of the full galaxy sample from DR7 is shown in Fig. \ref{fig:LAMOST_sn_distribution}.
We will discuss all selection criteria of the LAMOST spectra in more details in the next section and the sample with measured VDs in \S\ref{sec:veldisp_measurements}.

\subsection{Spectra selection for kinematic measurements}
\label{sec:selection}
Since we are interested in extracting the galaxy VD from LAMOST spectra, the main criteria to select high-quality spectra is the SNR. As anticipated, we have followed the prescription from the LAMOST data releases and used as first criteria the $SNR >10$ in $g$- or $i$-band.
%We have checked if lower SNR could provide robust estimates, but for this first velocity dispersion catalog, we finally decide to adopt a conservative cut and leave a more effective selection toward lower SNRs for future releases.
%Looking at the distribution of the SNR in $g$- or $i$-band in Fig. 2, we can see that if one wants to adopt the SNR$>10$ criteria suggested by the LAMOST pipeline, one would end up with a large fraction of discarded systems.
The LAMOST pipeline also provides the keyword  {\tt FIBERMASK}, which reports possible problems of the fiber: this is an integer value which is $0$ for fibers with no problem and larger than 0 for fibers with issues (not allotted fiber, bad centering, the too low flux in sky flat frames, bad arc solution, $>10\%$ pixels are bad on CCD, $>10$ pixels are saturated, whopping fiber, near a whopping fiber, extreme residuals). As we will discuss in \S\ref{sec:veldisp_measurements}, the pipeline used to measure VDs will also perform an independent assessment of the spectra quality, focused on the fitted wavelength window. %\st{features that are useful for the spectral fitting. For this reason, we will keep this keyword as a record, but mostly rely on the spectral quality selection we will perform.}
%Hence in the following we will also consider only spectra with ${\tt FIBERMASK}=0$.
%Finally, the quality of the VD estimates will be further assessed by other quality metrics related to the fitting procedure in \S\ref{sec:veldisp_measurements}.

Finally, the robustness of the VD estimates will also be checked against literature velocity dispersions from {SDSS-Data Release 7 (SDSS-DR7 in short) and BOSS} which have galaxies in common with LEGAS.\\

%In particular, among the above criteria, the one on the SNR threshold is looser with respect to the usual more conservative SNR thresholds adopted for velocity dispersion measurements \citep[e.g., ][suggest SNR>8]{Hopkins2013}. This is mainly due to the fact that this is an average SNR computed by the GAMA pipeline which does not necessary correlate with the SNR of the absorbtion line features we need to use for the kinematics. After some tests, we decided to keep SNR=1.5 as a safe threshold in order to select the sample of galaxies to run our procedure on. We leave a more conservative cut to be made {\it a posteriori} using the more realistic SNR computed from the bootstrap procedure we will adopt for the spectral fitting.

%high enough to perform reliable kinematical estimates. Usually the

After the aforementioned SNR cut, we ended up with a list of a collection of 148\,440 spectra, corresponding to 124\,214 galaxies, which have been plotted on the sky in Fig. \ref{fig:LAMOST_sky} and characterized in terms of their redshift and r-band magnitude distributions in Fig. \ref{fig:LAMOST_z} (blue colors).
%
%To perform velocity dispersion measurements, we have selected spectra with sufficient quality to allow reliable spectral template fitting (see \S\ref{sec:pPXF}). However
%We subsequently defined a quality index and checked a bootstrapping procedure in  to define a quality index for them
%which grades the robustness of the estimates (see \S\ref{sec:pixmask}).\\
%NRN I have noticed in the sentence before that the quality index results to be defined in a section before the bootstrapping. While in the sentence we seem to imply the other way around...
%We used this sample to perform the VD measurements, being as less conservative as possible with the spectrum selection criteria in order to maximise the sample of galaxies with velocity dispersion estimate. We will subsequently check the fit residuals, and define some further quality index based on the goodness of the fit (see \S\ref{sec:veldisp_measurements}).
%In Figure \ref{fig:GAMA_redshift_distribution} we show the distribution of the redshift of this selected sample together with the SDSS/BOSS galaxies observed in the GAMA fields. From the figure we can see that criteria above do not produce any particular selection effect as a function of the redshift and hence the selected sample is a good representative of the GAMA/AAT database.\\

However, all the criteria above do not guarantee that the spectra are suitable for the fitting procedure we want to perform, being the ratio of bad pixels/good pixels (on the main features considered for the spectrum fitting) the ultimate factor to assess the quality of the spectrum, which will be illustrated later (see \S\ref{sec:selection}). For this reason, after the criteria above have been used to preliminary select the spectra to be used for the fitting procedure, a further automatic selection is applied, which directly scans the spectra pixel by pixel to qualify them for the analysis.

\begin{figure}
\hspace{-0.5cm}
        \includegraphics[width=1.03\columnwidth]{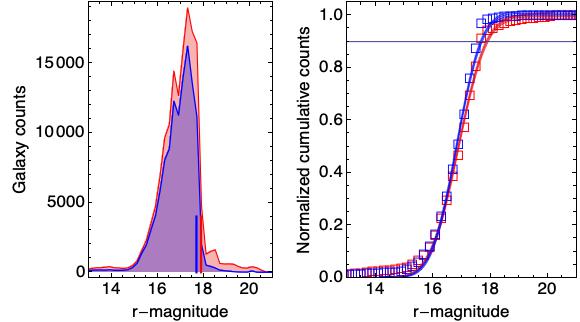}
    \caption{{\it Left}: Luminosity distribution of all sources labeled as galaxies (red), and the ones that have passed the SNR criteria (blue). The LAMOST selection function sharply selects galaxies around $m_r=18$. Short vertical lines show the 90\% relative completeness of the LAMOST sample (see right panel). {\it Right:} relative completeness of the LAMOST galaxies with respect to the total sample {all sources labeled as galaxies (red squares) and the ones that have passed the SNR criteria (blue squares)}. The horizontal line show the 90\% relative completeness of the LAMOST sample. Overplotted the best-fit error function as in Eq. \ref{eq:compl_erf}, with the $\pm1\sigma$ confidence curves from the best-fit parameters.}
    \label{fig:lamost_comp}
\end{figure}

We can now give a more quantitative estimate of the relative completeness as a function of the magnitude of the galaxy sample before and after the SNR cut. This is obtained by fitting the normalized cumulative counts of the two samples shown in Fig. \ref{fig:lamost_comp} with a standard error function model (see e.g. \citealt{Rykoff+015}).
\begin{equation}
{\rm comp}=(1/2)\left[Erf\left( \frac{m-m_{100}}{\sqrt{2w}} \right) \right],
\label{eq:compl_erf}
\end{equation}
where $m_{100}$ is the magnitude at which the completeness is 100\%, and $w$ is the (Gaussian) width of the rollover.
The normalized cumulative counts are obtained by the $r-$band magnitude distribution in Fig. \ref{fig:LAMOST_z} (bottom).
The magnitude at which the sample is 90\% complete with respect to the total sample has been extrapolated by the best fit function and turned out to be $mag_r=17.84\pm0.04$ and $mag_r=17.64\pm0.04$ (errors are nominal from the fit, see Fig. \ref{fig:lamost_comp}). These magnitudes give a reasonable measure of the completeness of the sample due to the LAMOST selection function that sharply selects galaxies around $m_r=18$.

\begin{figure*}
        \includegraphics[width=\columnwidth]{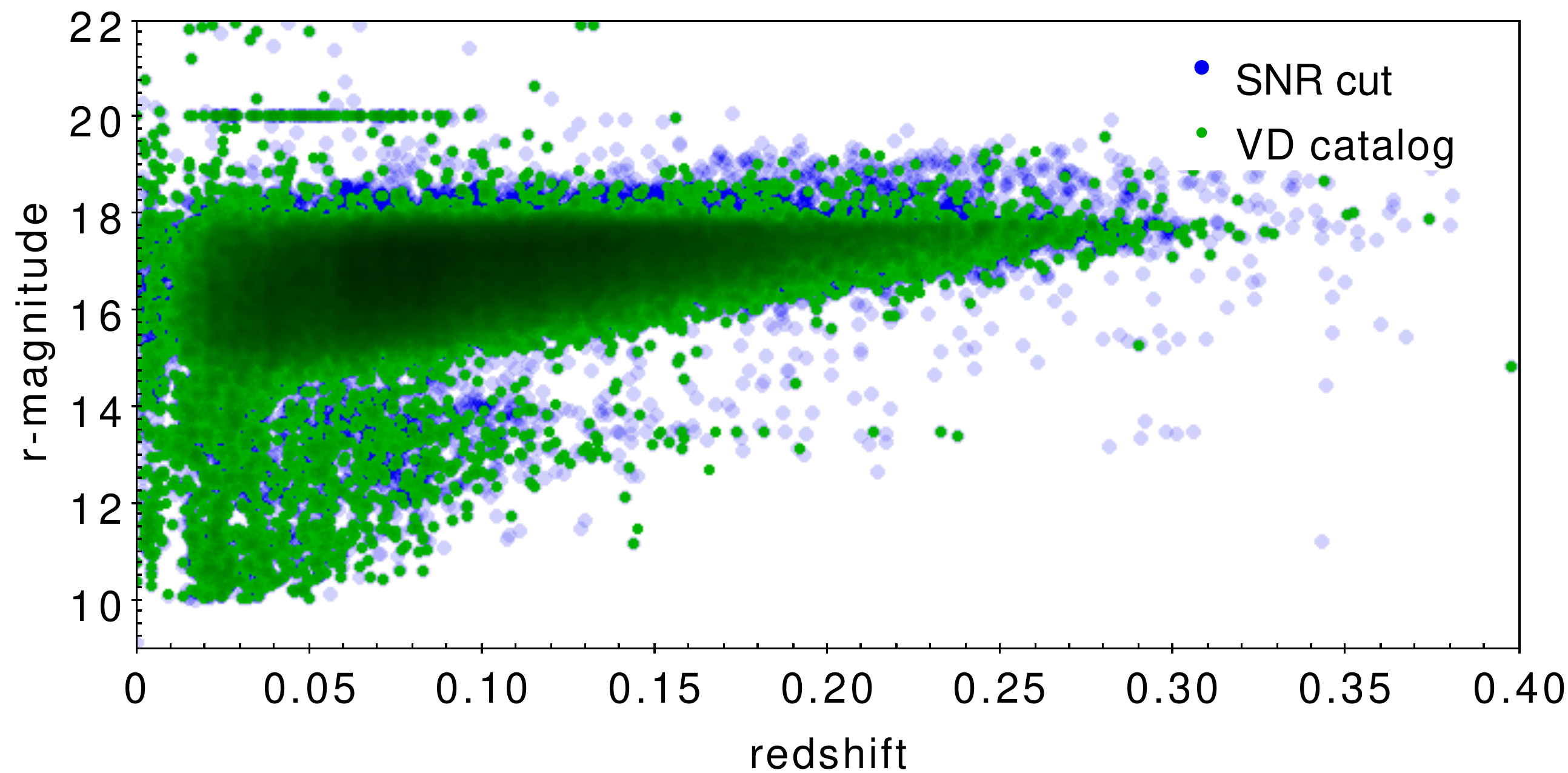}
        \includegraphics[width=\columnwidth]{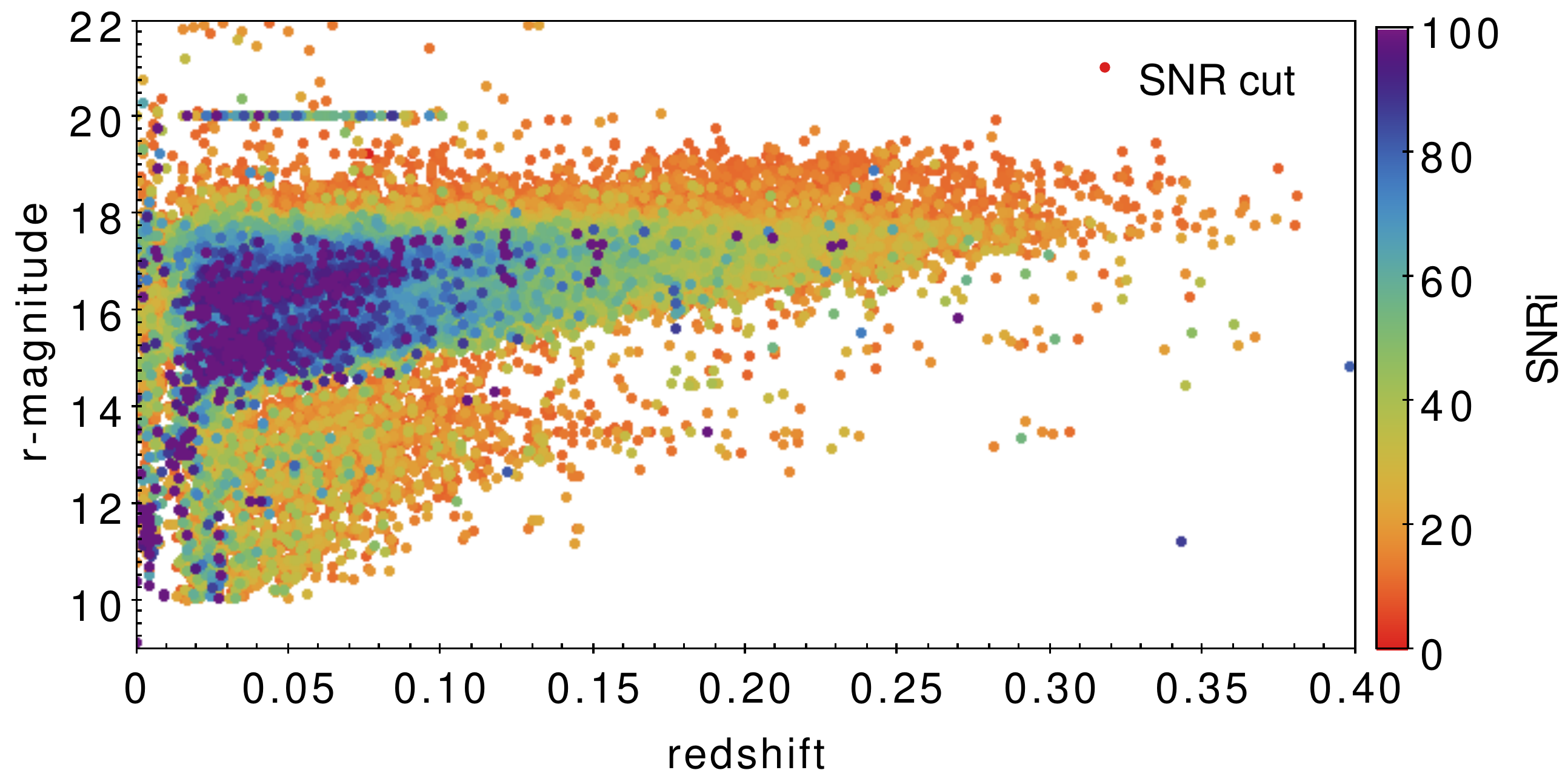}
    \caption{{\it Left}: $r-$band magnitude vs. redshift for the LAMOST galaxy spectra belonging to the sample selected for the velocity dispersion run through the SNR cut (blue) and the one for which we have obtained the velocity dispersion (green). {\it Right}: the qualified spectra for the velocity dispersion are color-coded according to their $i-$band SNR as indicative of their quality. This shows that the spectra SNR degrades mainly at magnitudes fainter than 18; however there are also a number of lower SNR at $mag_r<14$ and redshift $z>0.02$ that causes an incompleteness of the sample in this parameter space. Both panels include the galaxies with duplicated spectra.}

\label{fig:LAMOST_z_mag}
\end{figure*}

\section{Velocity Dispersion Measurements}
\label{sec:veldisp_measurements}
%NRN-0: to be integrated
As mentioned earlier the resolution of the LAMOST spectra $R\sim1800$, together with a sufficient SNR of the main spectral features that can be used for galaxy kinematics (i.e., CaII-K, CaII-H, H$\delta$, G-band and Mg), can be used to obtain kinematic information of the central galactic regions. Given this multitude of spectral features covered in the wavelength range, one can expect to reduce the typical errors of dispersion measurements to a few tens of $\rm km \, s^{-1}$, provided that these spectral features have a sufficient SNR (\citealt{Hopkins2013}). Although internal kinematics of galaxies was not a main driver of the LEGAS survey, VDs represent a very valuable added value for galaxy formation and DM studies, hence LAMOST can complement the SDSS/BOSS and GAMA samples, at redshifts $z<0.3$ in the NGC, and {\it provide wide coverage of galaxies in the SGC, where there is almost no similar data available} (see Fig. \ref{fig:LAMOST_sky}).

%This is possible, once one can take all source of systematics under control. In our case the largest uncertainties come from the correct definition of the spectral range where to apply the fit, the masking of the sky lines and gaseous emission lines; additionally pixels not corresponding to any flux measurement (i.e., \texttt{NaN}, zeroes, or extremely high values) can still be present in AAOmega spectra, even after the reduction. It is also important to take into account for the correct set-up in the use of fitting softwares, and we will later explain how we have dealt with possible internal biases of the pipeline used for performing our measurements. Normally, such bad features do not affect GAMA/SDSS spectra.\\
\subsection{Velocity dispersion pipeline}\label{sec:pPXF}
LAMOST spectra\footnote{The velocity dispersion pipeline works on 1D spectra provided by the LAMOST-DR7 database (http://dr7.lamost.org/). See \citealt{Luo+12_LAMOST} for details on the data reduction pipeline.} have been fitted using a fully automatic Python wrapper procedure %(D'Ago et al. 2020, in preparation) \gda{possiamo anche non citare nulla qui, perché stiamo applicando semplicemente ppxf}
which selects, prepares, and applies the Penalized Pixel-Fitting, \textsc{pPXF}, software (\citealt{Cappellari2017})
%\st{over a wide range of set-up conditions}
, in order to optimise the best-fit of galaxy spectra in the rest frame wavelength range, i.e. 3700-9900 \AA\ (see \S\ref{sec:spectra}).
%\st{\textsc{pPXF} is the state-of-the-art for fitting a whole galaxy spectrum which has been generally applied to small samples of galaxies observed with the same instrument and high SNR (see e.g. ATLAS3D, references).}
%As we will show later, it has been tested over mock spectra to estimate accuracy and precision and finally compared with SDSS/BOSS-DR14 central dispersion estimates over a common galaxy sample, which showed a very good agreement, within the statistical errors.

%\subsection{Penalized Pixel-Fitting}\label{sec:pPXF}
%The core of our procedure is based on the Penalized Pixel-Fitting \citep[\textsc{pPXF, }][]{Cappellari2017}, which we use to fit the galaxy spectra with broadened stellar templates.
\textsc{pPXF} works in pixels space, after having logarithmically rebinned the wavelength of both the observed spectrum and template library spectra. The $\chi^2$ minimisation is performed by making use of the Levenberg-Marquardt method. In the \textsc{pPXF} set-up, we make use of additive polynomials in order to reduce template mismatch and correct for imperfect sky subtraction and scattered light.\\
Since \textsc{pPXF} works on restframed spectra, the set-up of the software is made particularly easy by knowledge of galaxy redshifts. In order to apply \textsc{pPXF} to a large sample as LAMOST, we need to pay attention to the spectra selection and pre-processing (see \S\ref{sec:spectra-assess}).
%\textsc{pPXF} is the state-of-the-art for fitting a whole galaxy spectrum which has been generally applied to small samples of galaxies observed with the same instrument and high SNR (see e.g. ATLAS3D, references) in an interactive way.
%When dealing with large sample of galaxies, observed under different conditions of sky and in different nights, we have to deal with a large variety of spectra which show many different defects that cannot be handled interactively.
%Furthermore, if the SNR is not sufficiently high, the spectral fit with \textsc{pPXF} is not a trivial task. GAMA/AAT spectra have on average a declared low SNR (i.e., mean-SNR=5.29 and median-SNR=4.4, see Figure \ref{fig:GAMA_sn_distribution}) and are easily affected by significant noise in the main absorption lines which are critical for the template fitting. \\
A similar approach has been already applied to obtain galaxy internal kinematics from other instruments (see e.g. \citealt{Tortora+18_UCMGs}; \citealt{Scognamiglio+20_UCMGs}).

In the following we will illustrate all the main steps of the pipeline and the fitting strategies.

\subsubsection{Spectral range}
\label{sec:spectra-assess}
The first step of our analysis has been to
%In order to run \textsc{pPXF} on a large dataset, we need to
pre-process LAMOST spectra and obtain the most homogeneous set of spectra, thoroughly masked for bad features, and restframed. For this latter step, we can rely on robust redshift measurements from previous LAMOST releases and easily de-redshift the LAMOST spectra before applying \textsc{pPXF}.
%is then used in the different stages of our pipeline, as we will describe in details in the next section.
The ``high-quality'' LAMOST spectra (selected according to the criteria in \S\ref{sec:selection}) have been passed to the pipeline to be analysed. Given the $z-$distribution of LAMOST galaxies (mainly at $z<0.065$, see Fig. \ref{fig:LAMOST_z}), we decided to restrict our analysis to the restframe wavelength range: 3800-6000 \AA$\mbox{}$ in order to include only relevant absorption lines such as CaII-K, CaII-H, G-band, Mgb, and NaD.
%\st{The first wavelength range is selected for spectra with redshift $z > 0.1$, for which the CaII lines are less affected by the noise and can be more robustly used for the template fitting. The second one is selected for spectra having $z < 0.1$, for which the blue end of the spectrum is more affected by the noise.}
The adoption of a restricted (albeit still large) wavelength range has also dictated to reduce the computing time, which is mandatory for large datasets to limit the analysis to a few days on a personal desktop/laptop. To do that we needed to rest-frame the spectra and trim them to match one of the two selected wavelength ranges.

\subsubsection{Templates and rebinning}\label{templates_rebin}
%As done for other datasets, where the
%\st{same pipeline}
%a similar approach has been already succesfull applied (Tortora et al. 2018, Scognamiglio et al. 2020),
We used a selection of 40 MILES simple stellar population synthesis models (SSPs) from \citet{Vazdekis2010} as reference templates. {The choice of a limited number of templates allowed us to reduce the computational time, without affecting the accuracy of the VD estimates (see Appendix \ref{sec:app1} for a discussion).}
The selected SSPs have a unimodal IMF with logarithmic slope $\Gamma$=1.3, metallicity ranging between -1.71 and 0.22, and age ranging between 1 Gyr and 12 Gyr. %(we have checked that a wider library would not have improved the velocity dispersion measurements).\\
Following the \textsc{pPXF} prescriptions, template spectra are re-binned at the same velocity sampling of the LAMOST spectra. Subsequently, the template spectra are convolved with a Gaussian profile at the same instrumental resolution of 2.4 \AA\ \citep{2015MNRAS.448..822X} after having
%NRN: Peppe per favore puoi controllare cosa manca qui?
corrected the FWHM by the galaxy redshift (i.e.  FWHM$_{rest}$=FWHM$_{obs}/(1+z)$, as prescribed by \textsc{pPXF}).
%NRN: I'm not quite sure what the correction above means. GD: it is in PPXF manual. We need to restframe FWHM too.
\subsubsection{Pixel-masking}\label{sec:pixmask}
A pixel mask is needed to exclude outlying pixels and bad features from the fitting procedure. As a first step, the pipeline conservatively masks the wavelengths of typical gas emission lines, in order to avoid possible contamination by emission lines, like: H$\rm {\delta}$4101.74\AA, H$\rm {\gamma}$4340.47\AA, H$\rm {\beta}$4861.33\AA, $\rm [OIII]$4958.92\AA, $\rm [OIII]$5006.84\AA.
%Excluded pixels are recorded in a bad pixel mask.
We have checked that this conservative masking does not affect the measurement.\\
In addition to this brute force masking of typical emission lines, \textsc{pPXF} performs a preliminary fit and we take advantage of a sigma-clipping cut at 3-$\sigma$ level of the remaining hot pixel.
A spectrum having more than 50\% of bad pixels is automatically discarded from the analysis.\\
Finally, all spectra which pass this quality assessment are {restframed using the LAMOST catalogued redshifts and} processed with \textsc{pPXF}, by making use of an additive polynomial of the 4-th degree and letting the program to fit only two moments of the LOSVD {(i.e.  assuming a Gaussian distribution). Beside the redshift used to restframe the spectra, we used a $\sigma_{\rm start}=200$ kms$^{-1}$ as initial condition for all galaxies in the \textsc{pPXF} run\footnote{{The choice of the \textsc{pPXF} set-up, including the degree of the additive polynomials and the initial conditions, has been optimized though a series of tests on a limited number of spectra. For instance, we have checked that higher order polynomials did not produce any significant improvement in the accuracy of the \textsc{pPXF} best-fit. This is a standard approach in \textsc{pPXF} procedures. However one of the next improvement of the pipeline will be to perform bootstrap VD estimates letting all parameters to randomly varying within reasonable intervals and analyse the scatter of the \textsc{pPXF} inferences. A preliminary test is reported on Appendix \ref{sec:app1}, together with a more expanded discussion on the \textsc{pPXF} set-up}.}}.

\subsection{The LAMOST velocity dispersion catalog}\label{sec:results}
In this section, we describe the further criteria we have adopted to obtain a catalog of reliable velocity dispersion estimates and evaluate some completeness magnitude. These include some criteria on the $\chi^2$ of the best-fit model of the spectra and a range of acceptable values for galaxy-like VDs. We also use the information from multiple spectra of the same source to map the internal errors of the method. Finally, we describe the content of the delivered catalog.

\begin{figure*}
        \includegraphics[width=\columnwidth]{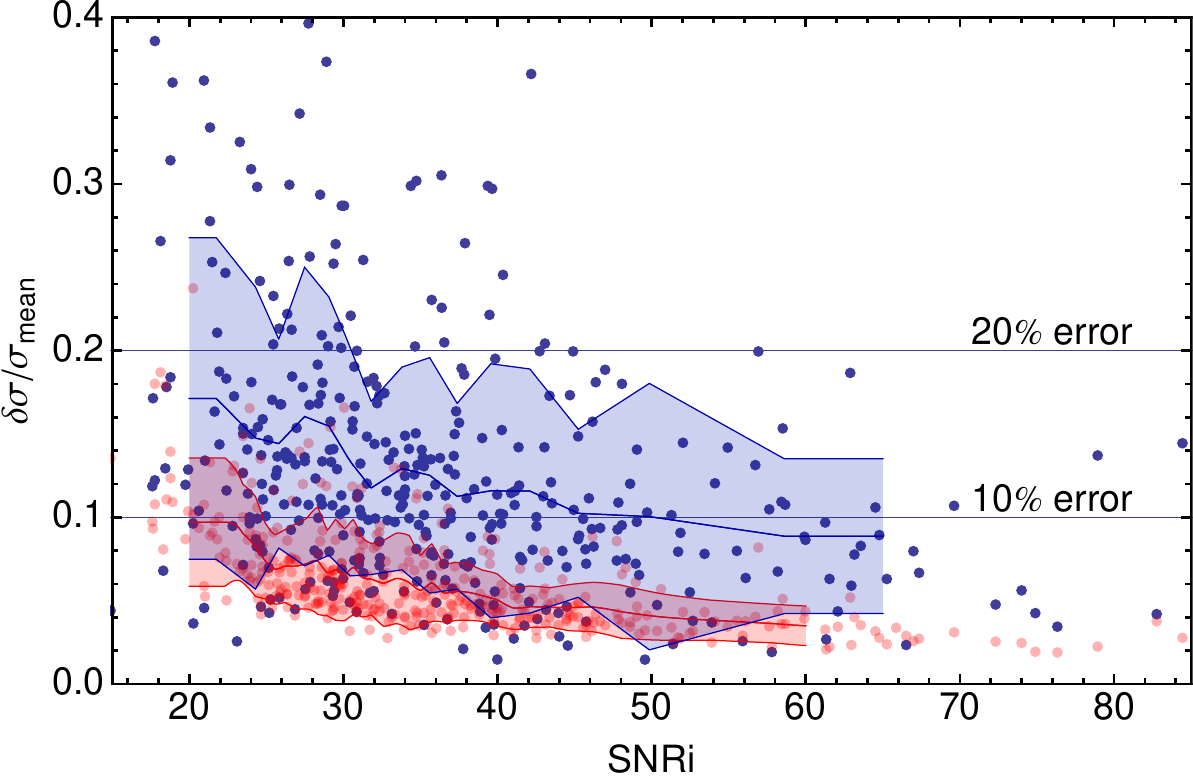}
        \includegraphics[width=\columnwidth]{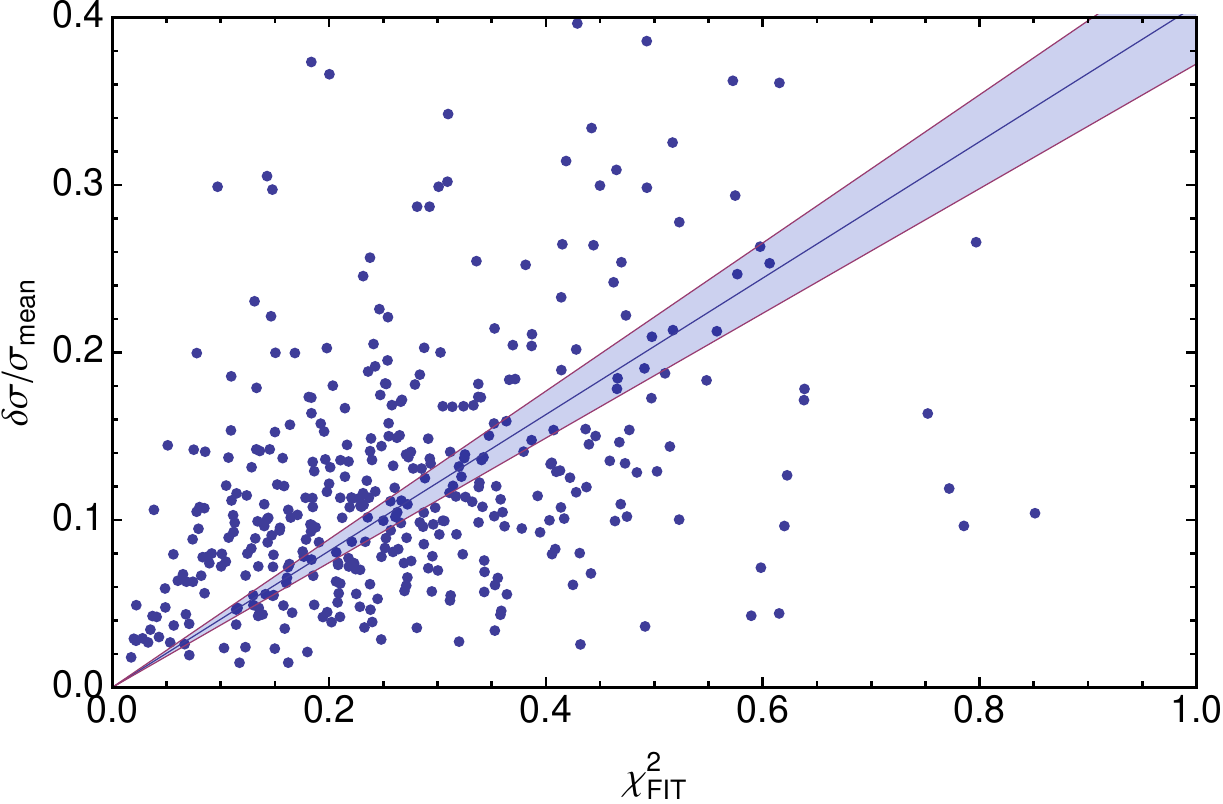}
    \caption{{\it Left}: statistical relative errors from repeated spectra of individual galaxies, $\delta\sigma/\sigma_{\rm mean}$ (blue dots, see text for a detailed description) and relative errors from \textsc{pPXF} (red dots) as a function of the $i-$band SNR as a proxy of the overal spectra SNR. Overplotted the shaded areas corresponding to a $\pm1\sigma$ scatter around a running mean. {\it Right}: $\delta\sigma/\sigma_{\rm mean}$ as a function of the best-fit $\chi^2$. Here there is a clear linear correlation which is quantified by the linear fit overplotted to the individual galaxy values (blue dots). The shaded area show the $3\sigma$ confidence interval of the best-fit parameters.
    }
\label{fig:rel_err}
\end{figure*}

\subsubsection{$\chi^2$ and velocity dispersion cuts}
\label{sec:catalog_cuts}
The \textsc{pPXF} run has returned the velocity dispersion for all spectra fed in the pipeline (i.e. 148\,440 entries). However, we have performed an \emph{a posteriori} quality check based on three main criteria:
\begin{enumerate}
    \item signal-to-the-residual ratio (SRR)>5;
    \item $\chi^2_{\rm FIT}$ < 1.1;
%    \item $\sigma_{\rm meas}$ > $\Delta\sigma_{\rm meas}$;
    \item 50 km/s $\leq\sigma_{\rm meas}\leq$ 450 km/s.
%NRN: Peppe per favore puoi aggiungere una riga per rispondere al dubbio sollevato da Crescenzo, per chiarire nel testo qui che le misure possono andare al di sotto dell'accuratezza nominale data dalla risoluzione? E anche controllare la questione half velocity scale che credo ho preso dalla descrizione del lavoro su GAMA?
\end{enumerate}
where SRR is the ratio between the median value of the flux and the \textit{rms} of the residuals of the fit retrieved with \textsc{pPXF}, and $\chi^2_{\rm FIT}$ is the reduced $\chi^2$ of the \textsc{pPXF} fit measured on the considered wavelength range. The lower velocity dispersion limit is adopted to cut  those galaxies with a measured sigma lower than 2/3 of the velocity scale of the instrument. By accepting these galaxies, we want to introduce in our final catalogue those measurements which are still higher than the MILES library resolution and which can be statistically significant for general purpose analyses without introducing biases due to a too stringent selection based on the sole velocity scale of the instrument. Similarly, we cut those galaxies with unrealistically high velocity dispersion ($>$450 km/s) considering the galaxy sample we are analysing\footnote{Typical stellar masses are of the order of $\log M_*/M_\odot<12$ (\citealt{Chabrier01} IMF), see \S\ref{sec:mass-sigma}.}
%, but still high enough not be considered a too stringent limit.\\

Using these criteria, we ended up with a sample of 98\,340 galaxies with measured velocity dispersions. These become 99\,515 if we use a looser cut on velocity dispersion of 35 km/s $<\sigma_{\rm meas}<$ 450 km/s.
%The lower value of this last selection is meant to cut away those galaxies with a measured sigma lower than half of the velocity scale of the instrument. By accepting these galaxies, we want to introduce in our final catalogue those measurements which are still higher than the MILES library resolution and which can be statistically significant for our subsequent analysis without introducing biases due to a too stringent selection based on the solely velocity scale of the instrument. By exploiting a similar argument, we cut those galaxies with unrealistically high velocity dispersion ($>$450 km/s) considering the sample we are analysing, but still high enough not be considered a too stringent limit.\\
%\st{while the higher threshold is meant to cut those galaxies with an unrealistically high velocity dispersion, considered the sample we are analysing. Such thresholds are meant to be non-stringent constraints in order not to introduce biases.}\\

To understand what are the galaxies missed after this last selection criteria, in Fig. \ref{fig:LAMOST_z_mag} (left panel) we show the $r-$band magnitude of the galaxy samples fed into the velocity dispersion pipeline (i.e. that qualifies because of the SNR criteria) and that survived the cut above after the velocity dispersion measurements. In the same figure (right panel) we also show the ingested galaxies color-coded by their SNR in $i-$band (as a proxy of their quality). From these plots, it is evident that the galaxies with velocity measurements tend to occupy the regions of the plot with higher SNR.
However, there is a region with lower SNR at $mag_r<14$ and redshift $z>0.02$ that produces a lower incompleteness of the velocity dispersion sample. From the figure, it is also trivially clear that the SNR increases at lower redshift and/or in the brightest galaxies.

\subsubsection{Repeated spectra: realistic relative errors and duplication cleaning}
\label{sec:repeated}
As mentioned earlier, the LAMOST database contains repeated spectra, i.e. multiple observations of the same sources. We have checked that in the sample of 98\,340 spectra, which survived after the criteria applied in \S\ref{sec:catalog_cuts}, there are 22\,194 which are repeated spectra (from 2 to 22 re-visit) of 9\,715 sources, while the remaining 76\,146 have a single spectrum available. Obviously, we need to define the best velocity dispersion estimate for the repeated sources among the ones produced by our pipeline, and clean all the duplication from the final catalog. We will address this point at the end of this section. Before proceeding to that, we notice that the repeated velocity dispersion measurements from independent spectra can allow a direct and quantitative evaluation of the overall statistical uncertainties of the full pipeline, as this is applied to different spectra of the same galaxy with different properties like SNR, noise, calibration etc.
To evaluate these statistical errors we have selected all galaxies with several spectra larger than 4\footnote{This is the best compromise to allow some variance for the individual galaxies and to keep a significant sample of systems to see the variation of the errors as a function of high quality VD parameters.} (a total of 402 systems) and estimated for them: 1) the mean VD ($\sigma_{\rm mean}$), 2) mean velocity dispersion errors, 3) mean $i-$band SNR, 4) mean $\chi^2_{\rm FIT}$ and their standard deviations. For each galaxy, we could define the statistical relative errors as $\delta\sigma/\sigma_{\rm mean}$, where $\delta\sigma$ is the standard deviation of the different spectra for the individual galaxy and $\sigma_{\rm mean}$ is defined as above. In Fig. \ref{fig:rel_err} we show the relative VD errors as a function of the SNR in the mean $i-$band SNR and the mean $\chi^2_{\rm FIT}$ obtained from the same repeated spectra for each galaxy. First, we notice that the statistical relative errors are larger than the nominal ones that one would obtain from the errors given by the \textsc{pPXF} output. These latter are given in the same Fig. \ref{fig:rel_err} (left panel) as red dots, which are obtained by the ratio of the mean VD errors from \textsc{pPXF} in the repeated estimates and $\sigma_{\rm mean}$. In particular, the mean relative errors are about half of the statistical relative errors at any given SNR value. Second, as expected the statistical relative errors (as well as the mean \textsc{pPXF} relative errors) are a decreasing function of the SNR and go from $0.16\pm0.12$ at SNR$_i=20$ to $0.10\pm0.08$ at SNR$_i=80$.

\begin{figure}
%   \centering
    \includegraphics[width=\columnwidth]{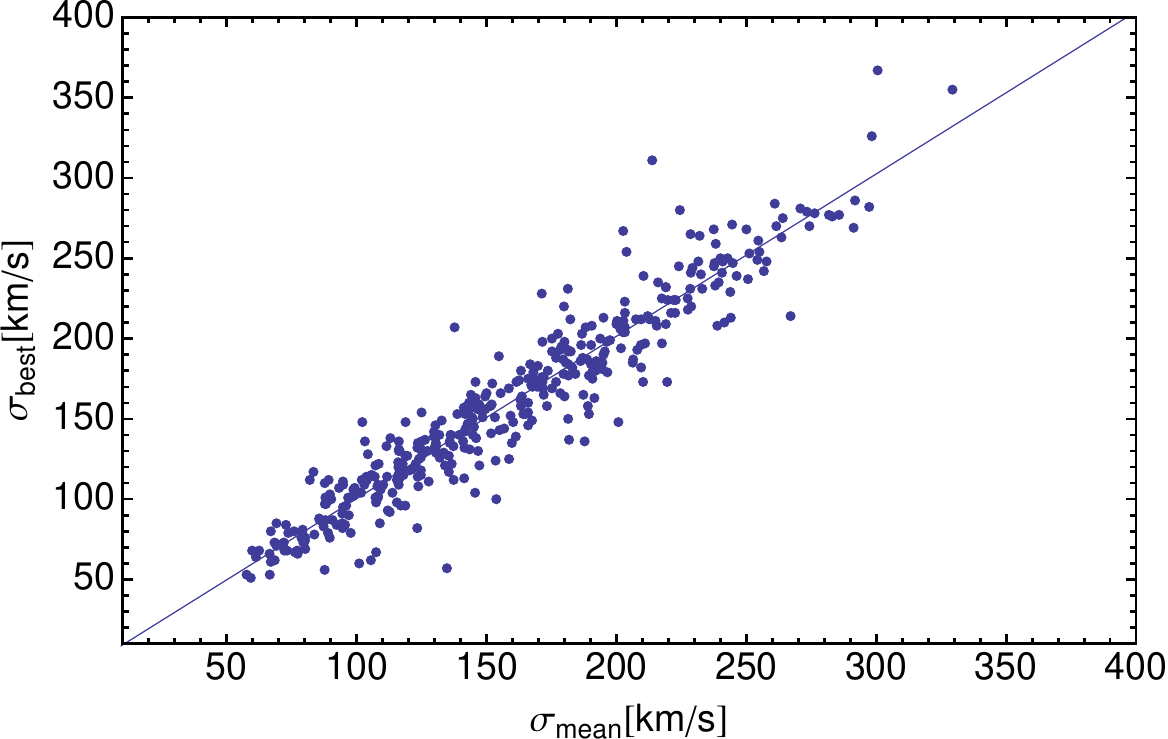}
    \caption{Mean VD from more than 4 repeated spectra, $\sigma_{\rm mean}$, vs. the best estimate, $\sigma_{\rm best}$, for each galaxy, given as the VD corresponding to the lowest $\chi^2_{\rm FIT}$ among the repeated measurements. Overplotted to the individual datapoints, the one-to-one line to show the perfect match between the two quantities.
    }
\label{fig:best_duplic}
\end{figure}

In the right panel of Fig. \ref{fig:rel_err} we show the statistical $\delta\sigma/\sigma_{\rm mean}$ as a function of the $\chi^2$ from the \textsc{pPXF} to the full spectra. Here we see that there is a clear correlation between the relative errors and the quality of the overall fit, i.e. the lower the $\chi^2_{\rm FIT}$ the lower the $\delta\sigma/\sigma_{\rm mean}$.
%This is quantified by
{To quantify this we have: 1) estimated a Spearman's rank correlation coefficient, $\rho=0.407\pm0.004$ (where the error is the standard deviation over 1000 jackknife extractions of the 2 parameters randomly cutting 1\% of the sample at every run), which corresponds to $>99\%$ correlation significance according to the Student's t-distribution ($dof=$n. data$-$2=400); 2)} performed the linear fit $y=Ax$ where $A=0.42\pm0.02$ {(hence significantly different from zero)}, which is also shown in the same panel with the shaded region corresponding to the $3\sigma$ confidence level. This tight correlation suggested us to use the lowest $\chi^2_{\rm FIT}$ to pick the best estimate, $\sigma_{\rm best}$, for each of the 9\,715 galaxies with multiple measurements. In Fig. \ref{fig:best_duplic} we plot $\sigma_{\rm best}$ with the $\sigma_{\rm mean}$ for each of the 402 galaxies as in Fig. \ref{fig:rel_err}: they are perfectly on the one-to-one relationship with a very small scatter{, defined as the standard deviation of the quantity $1-\sigma_{\rm mean}/\sigma_{\rm best}$), and found to be $\sim0.12$, consistently with average statistical errors on Fig. \ref{fig:rel_err} (left panel).} This means that the $\sigma_{\rm best}$ is a  good probe of the true velocity dispersion of these systems.
%\subsubsection{Final catalog without duplication}
%\label{sec:rel_err}
We have then obtained the $\sigma_{\rm best}$ for all the galaxies with multiple spectra (9\,715) which we added to the single spectra systems (76\,146) to obtain a final catalog of 85\,861.
We summarize all the samples discussed in the last sections and relative numbers in Table \ref{tab:tab1}.

\begin{table}
%   \begin{minipage}{120mm}
    \centering
    \begin{tabular}{l r r r r}
        \hline
                &All & Galaxies & Repeated & Single\\
        \hline
        DR7 Spectra &193\,883 & 163\,765 & 53\,128 & 140\,775\\
        SNR cut &148\,440 & 127\,093 & 36\,348 & 112\,092\\
        VD catalog  &98\,340 & 85\,861 & 22\,194 & 76\,146\\
        \hline
    \end{tabular}
    \caption{LAMOST VD Catalog Content. Reported the number of LAMOST spectra of the DR7 and the relative number of Galaxies, for the whole sample, the adopted SNR cut and the final catalog of VD measurements. Also reported the number of single and repeated spectra for each sample.}
    \label{tab:tab1}

%    \end{minipage}
\end{table}

\begin{figure*}
        \includegraphics[width=0.99\columnwidth]{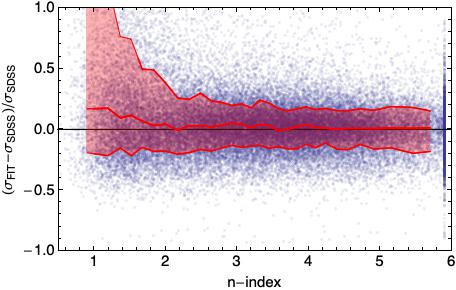}
        \includegraphics[width=\columnwidth]{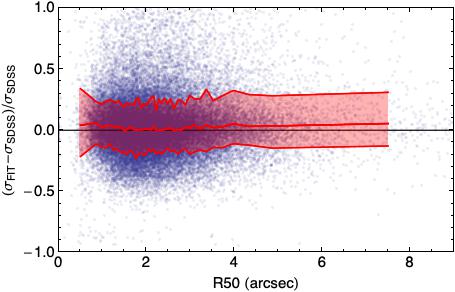}
        \includegraphics[width=0.99\columnwidth]{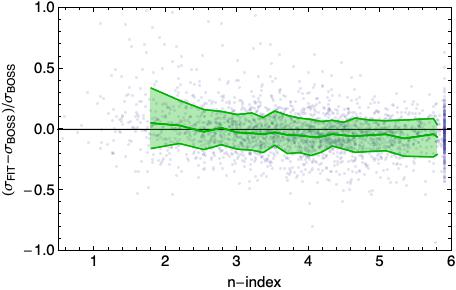}
        \includegraphics[width=\columnwidth]{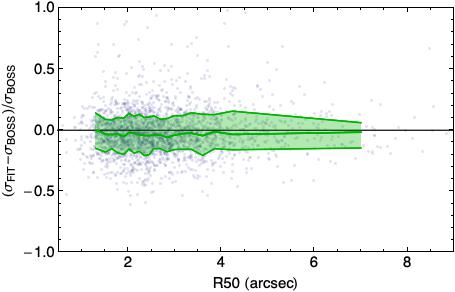}

        \caption{The relative deviation of the LAMOST measured VDs from \textsc{pPXF}, $\sigma_{\rm FIT}$, and SDSS estimates, $\sigma_{\rm SDSS}$ (top), or BOSS, $\sigma_{\rm {BOSS}}$ (bottom), as a function of $n-$index (left panel) and effective radius equivalent parameter, sersic$-R_{50}$, (right panel). The red shaded regions show the 16\% and 84\% quantiles, while the central solid line shows the median of the $(\sigma_{\rm FIT}-\sigma_{\rm SDSS})/\sigma_{\rm FIT}$ distribution in both panels.}
\label{fig:lamost_sdss_re_n-index}
\end{figure*}

\subsubsection{Final catalog content}
\label{sec:catalog_content}
We release the catalog of the sample of 85\,861 galaxies as an added value dataset of the LAMOST-DR7 data release. This is a first version of the velocity dispersion catalog that we plan to refine in future releases by: 1) a more accurate cleaning of the spectra quality throughout a more systematic triage based on spectra inspection (e.g. using machine learning techniques to assess the spectra quality); 2) bootstrapping the results by changing the \textsc{pPXF} setup and 3) randomly resampling the spectra {(see Appendix \ref{sec:app1} for more details)}.

For this first LAMOST-DR7 catalog we report the following information:
\begin{itemize}
    \item \textsc{specid}, the unique LAMOST spectrum ID;
    \item \textsc{ra}, right ascension (J2000) of the spectrum;
    \item \textsc{dec}, declination (J2000) of the spectrum;
    \item \textsc{z}, the redshift already measured by LAMOST pipeline;
    \item \textsc{$\Delta z$}, the redshift offset between LAMOST and \textsc{pPXF} derived measurement (see below);
%NRN: Peppe puoi aggiungere in fondo a questa lista la descrizione dei parameteri che non sono spiegati prima, come chiede Crescenzo nel suo commento?
    \item \textsc{$\Delta z\_err$}, the redshift offset error;
    \item \textsc{veldisp}, the velocity dispersion;
    \item \textsc{veldisp\_err}, the velocity dispersion error;
    \item \textsc{sng}, the LAMOST released $g-$band SNR;
    \item \textsc{snr}, the LAMOST released $r-$band SNR;
    \item \textsc{sni}, the LAMOST released $i-$band SNR;
    \item \textsc{srr\_fit}, the {\sl a posteriori} measured SRR on the fitted wavelength window;
    \item \textsc{chi2\_fit}, the $\chi^2/DOF$ measured on the fitted wavelength window.
\end{itemize}
The offset in redshift measurements $\Delta z$ arises from the fact that {\sc pPXF} just refines, by matching the best linear combination of templates with the observed spectrum, the radial velocity starting from an initial redshift guess (which in our case is the one measured by LAMOST by making use of PCA methods, \citealt{2018ApJS..234....5Y}). The mean offset between the {\sc pPXF} redshift measurements and the ones released by LAMOST is $10^{-4}$ and the mean error $\Delta z_{\rm err}$, arising from the combination of the two measurement errors, is $10^{-3}$. One reason for the offset is the different wavelength range adopted by the VD pipeline, while LAMOST pipeline uses all wavelength range (see \citealt{2019ApJS..240...10D}, for a detailed discussion).

\begin{figure*}
%   \centering
        \includegraphics[width=0.99\columnwidth]{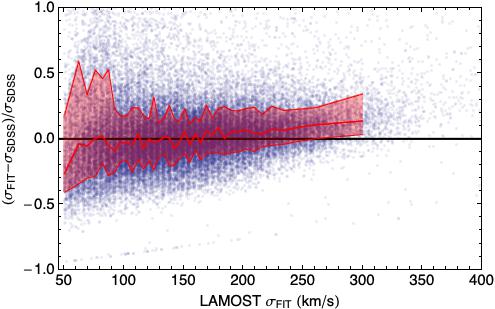}
        \includegraphics[width=\columnwidth]{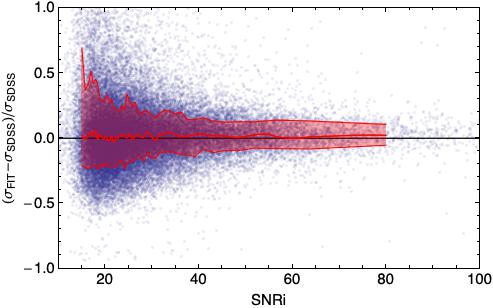}
        \includegraphics[width=0.99\columnwidth]{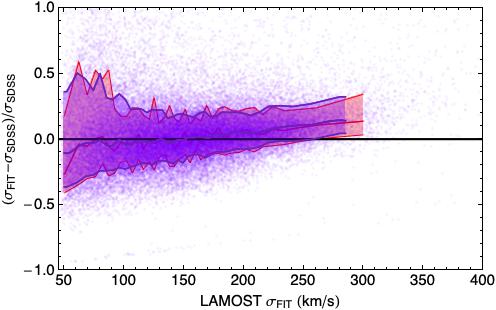}
        \includegraphics[width=\columnwidth]{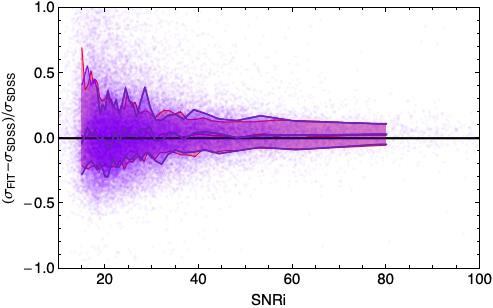}
    \caption{The relative deviation of the LAMOST measured VD from \textsc{pPXF}, $\sigma_{\rm FIT}$, and SDSS estimates, $\sigma_{\rm SDSS}$, from NYU-VAGC (\citealt{Blanton+05_NYU}) as a function of $\sigma_{\rm FIT}$ (left column) and the SNR in $i-$band (right column), for the original fiber aperture spectra (top row, in blue/red) and the aperture corrected VD values (bottom row, in purple, together with the corresponding uncorrected values as comparison). All shaded regions show the 16\% and 84\% quantiles, while the central solid line shows the median of the $(\sigma_{\rm FIT}-\sigma_{\rm SDSS})/\sigma_{\rm FIT}$ distribution.
    }
\label{fig:sigma_lamost_sdss}
\end{figure*}

\section{Comparison with SDSS and BOSS}\label{sec:tests}
We can now check the new LEGAS galaxy VD estimates against external catalogs of VDs measured on LEGAS targets. As discussed in \S\ref{sec:LEGAS}, LEGAS has a large overlap with SDSS (DR7) and BOSS (SDSS-DR12) in the NGC, and we can compare VD measurements on common objects. For the SGC, most of the information about LEGAS targets comes from the {Panoramic Survey Telescope and Rapid Response System (Pan-STARRS, \citealt{2016arXiv161205560C})} and no structural parameters, nor stellar masses have been associated with LEGAS galaxies.

\subsection{Comparison of the nominal estimates from fiber spectra}
\label{sec:no_ap_correction}
For SDSS we use the NYU-VAGC (\citealt{Blanton+05_NYU}) as a reference catalog as this provides additional photometric information, like the effective radius ($R_e$)\footnote{The best-fit Sérsic effective radii is from the $r-$band from the NYU-VAGC. These sizes are based on fits to azimuthally averaged light profiles and are equivalent to circularized effective radii.}, the Sérsic index ($n-$index, \citealt{1968adga.book.....S}) and the total Petrosian magnitude. In particular, since the $n-$index is distributed almost uniformly over the range of [0-6], i.e. LAMOST galaxies encompasses a wide range of galaxy types, we can test the consistency of the LAMOST estimates against SDSS as a function of the galaxy structural parameters. In general, to fairly compare independent measurements one should scale them at the same physical scale, e.g. the effective radius or some standard fraction of it (e.g. $R_e/8$, like usually used for fundamental plane analyses, see e.g. ref.). However, the difference between the LAMOST and SDSS fiber sizes is rather small, $R_{\rm Ap,lamost}=3.3''$ vs. $R_{\rm Ap,sdss}=3.0''$, so we expect this to have a minor impact on the measured kinematics from the spectra (see below).

The match of the LAMOST VD catalog (85\,861 entries) with the NYU-VAGC (2\,506\,754 entries) returned 66\,045 objects, although a quite significant number of the matched sources from SDSS have a VD=0 and cannot be used for our comparison. The final catalog with suitable VDs from SDSS is made of 50\,533 entries: we will refer to this as the {\tt LAMOST\_SDSS\_VD\_SERSIC} catalog, to indicate that for these systems we have both internal kinematics and Sersic structural parameters from SDSS.

In Fig. \ref{fig:lamost_sdss_re_n-index} we plot the relative deviation of the {\tt LAMOST\_SDSS\_VD\_SERSIC} estimates from the \textsc{pPXF} fit, $\sigma_{\rm FIT}$, and SDSS estimates, $\sigma_{\rm SDSS}$, defined as $\delta\sigma/\sigma=(\sigma_{\rm FIT}-\sigma_{\rm SDSS})/\sigma_{\rm SDSS}$, as a function of both $n-$index and effective radius. These parameters are important to figure how the different apertures can impact the estimates of the same galaxy from different instruments, as depending on the shape and scale of the light profile, the fiber sample different regions of the galaxy with different kinematical gradients (see below). As the $\delta\sigma/\sigma$ measures the excess of the LAMOST estimates with respect to SDSS, from Fig. \ref{fig:lamost_sdss_re_n-index} we observe that there are no overall systematic trends of LAMOST VDs with respect to SDSS for both structural parameters, but there is an increase of the scatter at the small $n-$index and a small overestimate for large $R_e$s, although within the statistical fluctuations. However, despite the tiny difference, the trend we see in the figure is compatible with one should expect by an aperture effect. Indeed, most of the differences rise for large sizes ($R_e>R_{\rm Ap}$), i.e. when the aperture is sensitive to the internal variation of the VD (typically within 1 $R_e$, see e.g. \citealt{2011MNRAS.414.2923K}), or for small $n-$index, i.e. we are in presence of late-type galaxies (LTGs hereafter, e.g. \citealt{2018MNRAS.477.2560B} found positive gradients of the VD for $n\lsim 2$). The aperture effect is more evident when comparing with BOSS, due to the much smaller fiber size with respect to LAMOST.

For BOSS we use the catalog from \citet[][1\,490\,820 entries]{2013MNRAS.431.1383T}, which provides stellar velocity dispersions, emission-line fluxes (both observed and de-reddened), equivalent widths and stellar masses (see below). The match with the LAMOST VD catalog as above includes 5\,553 galaxies only: we will refer to this as the {\tt LAMOST\_BOSS\_VD} catalog. For 2\,560 of these galaxies, we also obtained the Sérsic parameters from the NYU-VAGC for uniformity with the SDSS sample\footnote{We cannot find another publicly available catalog of structural parameters for the BOSS galaxies, hence we needed to restrict our comparison sample to the one matching with the NYU-VAGC}. We will use this latest catalog for the following comparisons (and dub this {\tt LAMOST\_BOSS\_VD\_SERSIC}).

This is  also shown in Fig. \ref{fig:lamost_sdss_re_n-index} (bottom), where we first notice that the scatter between the LAMOST VD estimates, $\sigma_{\rm FIT}$ and the BOSS estimates, $\sigma_{\rm BOSS}$, is smaller that the ones obtained for both the $n-$index (left panel) and $R_e$ (right panel) for SDSS.
%These features might have a physical origin: 1) the small $n-$indexes($<2.5$) generally characterize late-type systems and the aperture correction formula we used is calibrated on ETGs (see e.g. a discussion about the aperture correction dependence on the $n-$index in Bernardi et al. 2019); 2)
For SDSS, the relative errors are well within 20\% (except for $n-$index$\leq2.5$), while the one from BOSS is within 15\%\footnote{These are derived by the 0.16 and 0.84 quantiles in Fig. \ref{fig:lamost_sdss_re_n-index}, that are almost symmetric and representative of the standard deviation of the quantity.}. For SDSS the relative scatters are slightly affected by a non-negligible number of catastrophic systems, i.e. galaxies for which $|\delta\sigma/\sigma|>0.5$ ($\sim3400$ objects, corresponding to the 7\% of the {\tt LAMOST\_SDSS\_VD\_SERSIC}), in some cases also $>1.0$ ($\sim1500$ objects, or 3\%). For BOSS the sample is smaller and does not present any such catastrophic cases.

\begin{figure*}
    \centering
        \includegraphics[width=0.99\columnwidth]{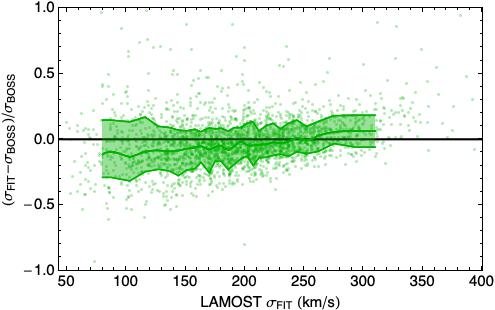}
        \includegraphics[width=\columnwidth]{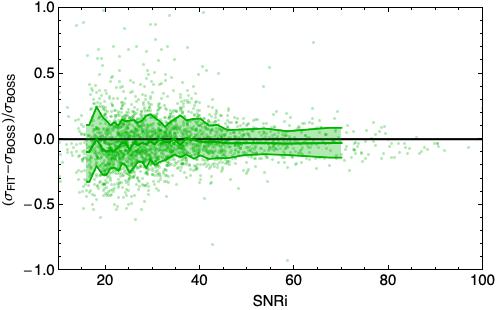}
        \includegraphics[width=0.99\columnwidth]{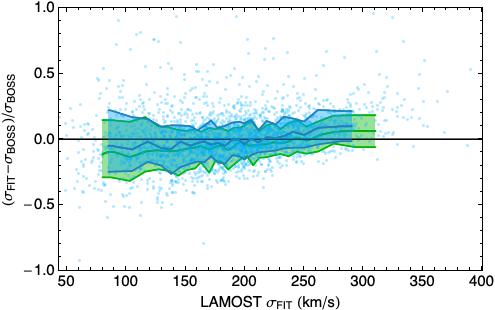}
        \includegraphics[width=\columnwidth]{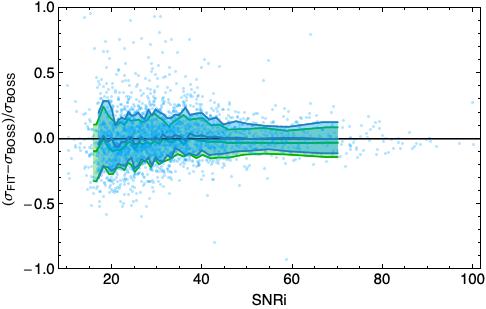}
    \caption{The relative deviation of the LAMOST measured VD from \textsc{pPXF}, $\sigma_{\rm FIT}$, and BOSS estimates, $\sigma_{\rm BOSS}$, from \citet{2013MNRAS.431.1383T} as a function of $\sigma_{\rm FIT}$ (left column) and the SNR in $i-$band (right column), for the original fiber aperture spectra (top row, in green) and the aperture corrected VD values (bottom row, in light blue, together with the corresponding uncorrected values as comparison). All shaded regions show the 16\% and 84\% quantiles, while the central solid line shows the median of the $(\sigma_{\rm FIT}-\sigma_{\rm SDSS})/\sigma_{\rm FIT}$ distribution.
    }
\label{fig:sigma_lamost_boss}
\end{figure*}

A second difference between SDSS and BOSS is that the latter, differently from SDSS for which we have discussed a lack of relevant systematics, show systematically smaller VDs with respect to BOSS (again with an inverse behaviour only for the lower $n-$index but smaller than the one observed for SDSS -- see the upper panel in Fig. \ref{fig:lamost_sdss_re_n-index}, left column). This is fully compatible with an aperture effect as the innermost regions sampled by BOSS are more biased toward the maximum of the radial profile of the VD and thus higher than the dispersion measured over a larger aperture, integrating over an area with smaller VD. Again the opposite trend in the lower $n-$index ($<2.5$) is due to some positive VD gradients (see above and \citealt{2018MNRAS.477.2560B}). In general the LAMOST estimates are about $5-10\%$ smaller than BOSS (e.g. looking at the $\delta\sigma/\sigma$ vs both $n-$index and $R_e$ in Fig. \ref{fig:lamost_sdss_re_n-index} bottom), with a mild decreasing trend with the $n-$index and almost no trend with $R_e$. In the next section, we discuss in more detail the comparison of the LAMOST estimates and the ones from SDSS and BOSS, after having taken the effect of the aperture correction into account. {\it These aperture corrected values will be also reported in the final velocity dispersion catalog attached to this paper}.

{To conclude this section on the direct comparison from fiber estimates against SDSS/BOSS, we want to remark that we are not considering other possible {sources of scattering}: e.g. the effects of site seeing and fiber positioning. While seeing is worth {exploring}, especially if one wants to use VD measurements for dynamical masses, the fiber positioning is harder to control. One way to check this is to consider whether the larger scattered values are correlated with the difference in SNR between SDSS and LAMOST (as one should expect in case of bad positioning). We did not find any particular trend, hence we concluded that fiber positioning should have a minor impact, at least over large statistical samples.}

\subsection{Comparison of the estimates after aperture correction}
\label{sec:ap_correction}
In order to evaluate the possible effect of the fiber sizes on the estimates from the different samples we have re-computed the VDs at the same physical size, the effective radius, also defined as the half-light radius.
In particular, we have used the formula by \cite{Cappellari+06} reported here below:
\begin{equation}
%\log(\sigma_{\rm R}/\sigma_{\rm R_e})=-\gamma(n, R, R_e)\log{R/R_e}
\sigma_{\rm R_e}=\sigma_{\rm R}(R/R_e)^{0.066},
\label{eq:ape_corr}
\end{equation}
%The match of the LAMOST VD catalog (85\,327 entries) with the NYU-VAGC (2\,506\,754 entries) returned 65\,606 objects, although a quite significant number of the matched sources from SDSS have a VD=0 and cannot be used for our comparison. The final catalog with suitable VDs from SDSS is made of 50\,201 entries: we will refer to this as the {\tt LAMOST\_SDSS\_VD} catalog.
although this has been  optimized for ETGs and it should be more accurate for $n-$index$>2.5$.
We decided to use this simple formula because we have seen in Fig. \ref{fig:lamost_sdss_re_n-index} that the $\delta\sigma/\sigma$ does not significantly depend on the $n-$index, hence one can ignore in the correction the effect of the light profile (although it might matter, see e.g. \citealt{2018MNRAS.477.2560B}) and consider only the effect of the ratio between the fiber aperture and the radius at which one wants to correct the $\sigma$. Hence despite it represents only an approximation, Eq. \ref{eq:ape_corr} has become a standard for this operation (e.g. \citealt{tortora2009}; \citealt{2011ApJ...737L..31B}; \citealt{Beifiori+14}; \citealt{Tortora+18_KiDS_DMevol}; \citealt{2020arXiv200201940S}).

In Fig. \ref{fig:sigma_lamost_sdss} we show the $\delta\sigma/\sigma$ before (top panels) and after (bottom panels) of the LAMOST estimated vs. SDSS-DR7 (in {\tt LAMOST\_SDSS\_VD\_SERSIC}), having applied the aperture correction for both $\sigma_{\rm FIT}$ and $\sigma_{\rm SDSS}$, as a function of the LAMOST VD estimates (left panels) and the sSNR in $i-$band (right panels). In the bottom panels we report the running mean also of the same quantities before the correction for direct comparison. The overall $\delta\sigma/\sigma$ before the correction shows a general agreement among the estimates although some deviation seems to emerge at higher $\sigma_{\rm FIT}$ ($>$200 km/s), while they are fully consistent to each other when plotted against the $SNR_i$. The situation does not substantially change after the aperture corrections, which indicates that the fiber diameters ar close enough between LAMOST and SDSS that the aperture correction is statistically indistinguishable from the uncorrected case. Finally the $\delta\sigma/\sigma$ also gives an independent estimate of the statistical errors of the estimates and, consistently with the results from repeated spectra, the scatter in the $\delta\sigma/\sigma$ is a strong function of the $SNR_i$ and goes from the 15\% relative error at $SNR_i>50$ to $\sim$20\% at $SNR_i=30-40$, while the scatter in the $\delta\sigma/\sigma$ is always $<$25\% for $\sigma_{\rm FIT}>100$ km/s, being the scatter quite larger at lower $\sigma_{\rm FIT}$.

In Fig. \ref{fig:sigma_lamost_boss} we show the same diagnostic as in Fig. \ref{fig:sigma_lamost_sdss}, but for BOSS galaxies. We can see here some trends quite similar to the ones in the $\delta\sigma/\sigma$ from SDSS, but with a smaller scatter (generally $<15\%$). In particular we notice a small but constant systematic negative offset ($\lsim5\%$) for the VD without the aperture correction, seen especially in the plot against $SNR_i$. This is a consequence of the larger differences in the fiber sizes which causes a significant aperture effect. However, as anticipated, the aperture correction renormalizes the estimates and re-aligns the overall trends of the aperture corrected $\delta\sigma/\sigma$ to the behavious found in SDSS. Therefore, using the aperture correction, we could consistently derive VD estimates that look clearly unbiased when plotted against the $SNR_i$ (see bottom right panel), but that still shows some (increasing) trend with the $\sigma_{\rm FIT}$, although they are all consistent for $\sigma_{\rm FIT}\lsim250$ km/s, but with a wider consistency with respect to the uncorrected estimates
%between LAMOST and BOSS
, both in terms of scattering and
%also consistency of the
median values.

\section{The Mass--$\sigma$ relation}
\label{sec:mass-sigma}

Galaxies are characterized by striking regularities in their properties, as the correlation between size and mass (e.g. \citealt{shen2003}; \citealt{HB09_curv}; \citealt{Roy+18}), or size and surface brightness (e.g. \citealt{Kormendy1977,Kormendy+09}; \citealt{Capaccioli+92a}). Then, there exist specific relationships for early-type galaxies, as the Fundamental Plane (e.g., \citealt{Djorgovski1987}; \citealt{Dressler+87}; \citealt{HB09_FP}) and the Faber-Jackson relation (\citealt{FJ76}) or those for late-type systems, as the Tully-Fisher relation (e.g. \citealt{TF77}; \citealt{Lelli+16_Tully-Fisher}). The study of these correlations, their evolution with redshift and the comparison with theoretical predictions provide crucial information about the galaxy formation and evolution processes, as demonstrated by the analysis of size and velocity dispersion evolution in massive early-type galaxies (e.g. \citealt{Fan+08}; \citealt{Cenarro_Trujillo09}; \citealt{Hopkins+09_DELGN_IV}; \citealt{Tortora+14_DMevol, Tortora+18_KiDS_DMevol}; \citealt{Roy+18}).

{
\begin{figure}
\hspace{-0.5cm}
        \includegraphics[width=1.05\columnwidth]{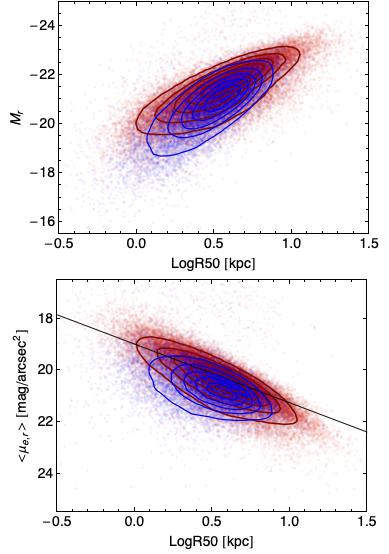}
    \caption{
    {\it Top}: the absolute luminosity$-$size relation of the LAMOST galaxies with available structural parameters.
    %The $M_r$  is obtained from the LAMOST redshift and $k-$correction is applied as described in the text.
    Galaxies are classified as ETGs (red), and LTGs (blue) according to their $n-$index. Isodensity contours in the parameter space are overplotted with the same color code. {\it Bottom:} the mean $\langle\mu_{e,r}\rangle$$-$$R_e$ relation of the same galaxy samples (ETGs, red points, and LTGs, blue points). Isodensity contours in the parameter space are overplotted with the same color code. The black solid line is the best fit to the $\langle\mu_{e,r}\rangle$$-$$R_e$ ($r-$band) relation of a local SDSS sample of ETGs from \citet{2012MNRAS.421.2277L}.
    %, vertically scaled by the median $r-i$ ($=-0.41$) color of the LAMOST galaxies. See text for details.
    }
    \label{fig:scal_rel}
\end{figure}
}

\begin{figure}
\hspace{-0.3cm}
        \includegraphics[width=1.02\columnwidth]{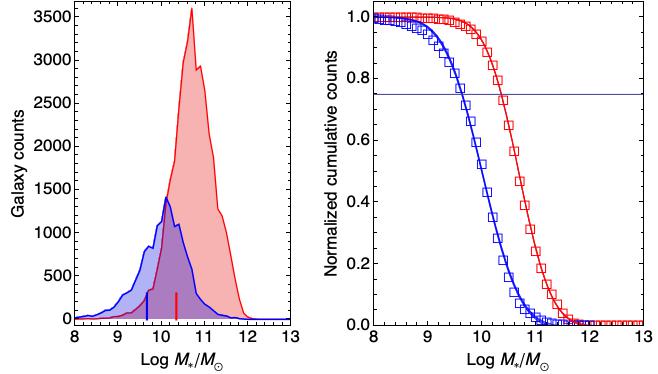}
    \caption{
    {\it Left}: Mass distribution of galaxies classified as ETGs (red), and LTGs (blue). They both show a pseudo-normal distribution. Short vertical lines show the 75\% relative completeness of the LAMOST sample (see right panel). {\it Right:} relative completeness of the LAMOST galaxies with respect to the total sample  {for the ETG sample (red squares) and the LTG sample (blue squares), as defined according to their $n-$index, see text}. The horizontal line shows the 75\% relative completeness corresponding to a mass limit where 50\% of the galaxies are lost with respect to the peak. Overplotted the best-fit error function, with the $\pm1\sigma$ confidence curves from the best-fit parameters (see discussion in the text)}
    \label{fig:mass_comp}
\end{figure}

Therefore, as a demonstration of the scientific validity of the dataset produces from LAMOST-DR7, we derive in this section the Mass-$\sigma$ relation for the LAMOST-DR7 galaxies and compare the results with independent analyses
%based on SDSS
{(e.g. \citealt{2011A&A...534A..61N}, \citealt{Tortora+14_DMevol})}. The Mass--$\sigma$ relation is a fundamental scaling relation for galaxies and it has been demonstrated to vary as a function of the galaxy properties (e.g. \citealt{Graves+09}) and as a function of redshift (e.g. \citealt{2011A&A...534A..61N}). We stress here that the following analysis, based on our newly derived VDs of LAMOST galaxies, is aimed at deriving the general trend of the relation to validate the dataset rather than fully tackle the physics behind the relation{, including the variation with redshift}. This will be the content of future analyses.

To derive the Mass--$\sigma$ relation for the LAMOST sample We have combined both the {\tt LAMOST\_SDSS\_VD\_SERSIC} and {\tt LAMOST\_BOSS\_VD\_SERSIC} catalogs for which we have 1) the aperture corrected VD estimates in \S\ref{sec:ap_correction} and 2) updated stellar masses, using star forming models according to \citet{2009MNRAS.394L.107M}, are provided by the SDSS Sky Server ({\tt stellarMassStarformingPort}\footnote{Direct download from this URL: https://www.sdss.org/dr12/spectro/galaxy\_portsmouth/ .}).
The match of the VD catalog with the stellar mass catalog has given 53\,308 entries.
We used these stellar mass models because a large part of our sample is made of LTGs as we could classify according to the $n-$index. Indeed, ETGs are roughly separated from LTGs, assuming $n>2.5$ for the former and $n \leq 2.5$ for the latter (\citealt{Roy+18}). Using this criterion we have found 37\,494 ETGs and 15\,814 LTGs.
{To illustrate the photometric properties of these galaxy samples, in Fig. \ref{fig:scal_rel} we show two classical scaling relations: the magnitude--size relation and the $\langle\mu_{e,r}\rangle$$-$$R_e$ relation (see also \citealt{2019PASA...36...35G} for a compelling review).
As the LAMOST galaxies cover a fair range of redshift (see Fig. \ref{fig:LAMOST_z}), absolute magnitudes in $r-$band have been obtained by applying a $k-$correction to the $r-$band apparent magnitudes\footnote{We used the ``K-corrections calculator'' service available at http://kcor.sai.msu.ru/}. We have grouped the galaxies in 5 redshift bins (from 0.1 to 0.5 in steps of $\Delta z=$0.1) and used the mean reshift of the bin and the individual galaxy $g-r$ colours to obtain the $k_r(z, g-r)$. Then these values have been used to derive the absolute magnitude in $r-$band defined as $M_r=m_r-5 \,{\rm Log} \,D_L/10pc - k_r$, where $D_L$ is the luminosity distance derived by the LAMOST redshift estimates. We have also applied the same $k_r-$correction to the mean surface brightness, defined as  $\langle\mu_{e,r}\rangle=(m_r-k_r)+2.5 \, {\rm Log} \, 2\pi +5\, {\rm Log}\, R_e$.

Albeit qualitative, the plots show regular trends of the ETGs which lie on a sequence that is different than the one of LTGs in both diagrams (see e.g.  \citealt{2009MNRAS.393.1531G}). Galaxies selected as LTGs are generally fainter and possess lower mean surface brightness in their centers (\citealt{2002MNRAS.332..422P}, see also \citealt{2020MNRAS.493.1686L}). The trends of LAMOST galaxies, though, incorporate some variation of the relations due to the variation with redshift of galaxy properties ($M_r$, $\langle\mu_{e,r}\rangle$, $R_e$). A full discussion of the dependence on redshift of these relations and its physical interpretation is beyond the purpose of this analysis, where we want to focus on the Mass--size relation, and it will be addressed in a forthcoming paper. However, in  the bottom panel of the figure, we overplot the best fit to the $\langle\mu_{e}\rangle$$-$$R_e$ ($r-$band) relation of a local SDSS sample of ETGs from \citet{2012MNRAS.421.2277L}.
%, vertically scaled by the median $r-i$ color (-0.41) of the LAMOST galaxies.
This shows an overall agreement with the LAMOST galaxies' relation, although their selection of ETGs is slightly more stringent than ours ($n-index=4$).
%where most of the low-redshift ETGs are located.

\begin{figure}

\hspace{-1cm}
        \includegraphics[width=1.01\columnwidth]{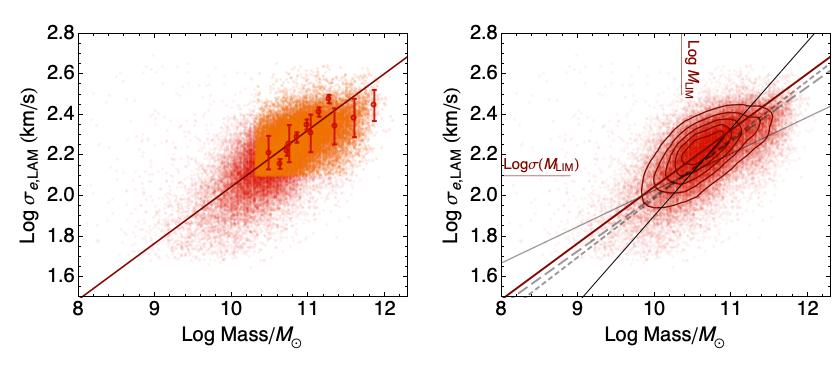}
        \includegraphics[width=1.01\columnwidth]{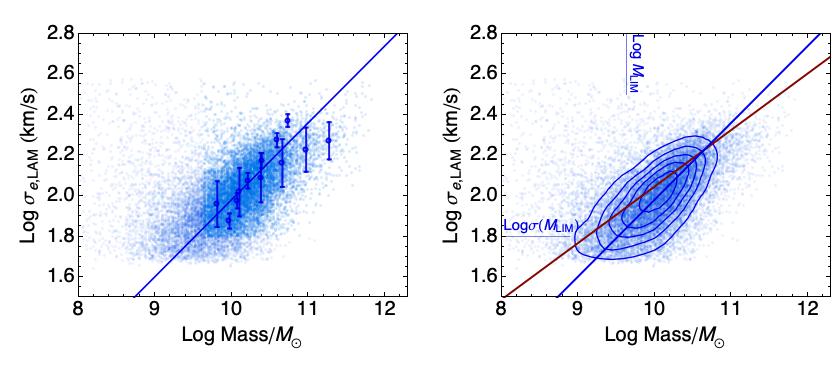}
        \caption{Mass--$\sigma$ relation from LAMOST galaxies, separated in ETGs (upper panels, red dots) and LTGs (lower panels blue dots). Velocity dispersion values are aperture corrected to match the value at the effective radius as in \S\ref{sec:selection}. Left panels: points with error bars are median and median deviation respectively of the binned data points, along the X-axis and the Y-axis. Solid lines (red on the top and blue on the bottom) are the weighted linear fit to the median values (see text for details) with the best fit values given in Table \ref{tab:tab2}. The lighter colored data points {(orange in the top panel and light blue in the bottom panel)} show the data above the completeness lines (see text for details and right panels). Right panels: density contours in the Mass--$\sigma$ space of the ETGs (top) and LTG (bottom). In both panels we report the best fit obtained from the median values, as in the left panel. For ETGs we also report literature relations from \citet{2011A&A...534A..61N} (black line) and a collection of best fit relations from \citet{Tortora+14_DMevol} (gray lines, see text for the full description). }
\label{Fig:mass_sigma1}
\end{figure}

Moving to the Mass--$\sigma$ relation, in} order to derive an unbiased trend
%of the Mass--$\sigma$ relation,
 we need to estimate a limiting mass below which we start loosing a significant number of galaxies with respect to the peak value. Following the procedure adopted in \S\ref{sec:selection}, we have used Eq. \ref{eq:compl_erf} to evaluate the mass at which the number of galaxies decreases at about 50\% from the peak. This is obtained by finding the mass at which the cumulative mass function shown in Fig. \ref{fig:mass_comp} equals 0.75. We have fitted the distributions of the ETGs and LTGs with the function 1-comp${x}$ as in Eq. \ref{eq:compl_erf}, where now the variable is the decreasing mass, and obtained $\log M_*/M_\odot=10.35\pm0.01$ and $\log M_*/M_\odot=9.63\pm0.01$ for ETG sand LTGs respectively.

\begin{table}
%   \begin{minipage}{120mm}
    \centering
    \begin{tabular}{c c c c}
        \hline
                \multicolumn{4}{c}{$\log \sigma_e=a + b \log M_*/M_\odot$}\\
        \hline
        Type & $a$ & $b$ & significance\\
        \hline
        ETGs& $-0.7\pm0.6$ & $0.28\pm0.06$ & $1\sigma$\\
        LTGs& $-1.8\pm0.7$ & $0.38\pm0.07$ & $1\sigma$\\
        \hline
    \end{tabular}
    \caption{Weighted linear fit to the Mass-$\sigma$ relation, according to the Eq. \ref{eq:mass_sigma}}
    \label{tab:tab2}
\end {table}

In Fig. \ref{Fig:mass_sigma1} we show the Mass, $\log M_{\rm *}/M_\odot$, vs. the
%aperture corrected
LAMOST velocity dispersion values corrected to
%the logarithm of
the effective radius aperture, $\log \sigma_{\rm e,~LAM}$, for the two galaxy types, ETGs (in red) and the LTGs (in blue). We also show the best fit linear relation to the galaxies above the limiting mass, $M_{\rm lim}$, and the average VD value ($\log \sigma(M_{\rm lim})$), i.e. the mean velocity dispersion that we have estimated for 400 galaxies in a bin around $M_{\rm lim}$. These are both indicated in the graph in order to evaluate the range adopted for the linear fit for ETGs and LTGs.

In order to mitigate biases in the linear fit due to the incompleteness both in mass and sigma, we have performed a linear fit over the median values we have derived binning the data along the X-axis and Y-axis, as indicated by the two groups of data with error bars (MAD/0.6745) in Fig. \ref{Fig:mass_sigma1} (left panels for ETGs and LTGs, on top and bottom respectively). The linear formula adopted is

\begin{equation}
\log \sigma_e=a + b \log{M_*/M_\odot}
\label{eq:mass_sigma}
\end{equation}

and the best fit parameters are reported in Table \ref{tab:tab2}, where we have used the inverse of the errors on the median points as weights. From the values of the slope ($b$) we see that the two samples are barely consistent to each other within $1\sigma$. We notice though that our fitting procedure is rather conservative, as it incorporates in its scatter the systematic of the trend in both variables.

%The median values are linearly fit  (with weights given as the inverse of the Y-axis errorbars): red is the best fit for the ETGs and the blue one for the LTGs.

We can compare these results with previous findings from local samples. In Fig. \ref{Fig:mass_sigma1} we show the relation found for different samples: 1) a volume limited sample of ETGs from the NYU-VAGC from \citet[N+11, solid black line]{2011A&A...534A..61N}, for which masses have been estimated from galaxy colours using the relation from \citet{2003ApJS..149..289B}\footnote{This conversion is based on a \citet{1955ApJ...121..161S} IMF (while we use \citealt{2001MNRAS.322..231K} IMF), hence we have re-scaled their relation to account for the 0.4 dex factor between the two mass estimates}; 2) a collection of Mass-$\sigma$ relations from \citet[T+14, see their Table 1]{Tortora+14_DMevol}, which has analysed data from different local samples from \citet[T+09, long dashed gray]{tortora2009}, \citet[Atlas3D, short dashed gray]{Cappellari+13_ATLAS3D_XV,Cappellari+13_ATLAS3D_XX} and \citet[SPIDER sample, solid gray]{SPIDER-VI}. We see that our ETG estimates are in a substantial agreement with the T+09 and ATLAS3D samples, in particular they show a very similar slope ($0.27\pm0.01$, $0.29\pm0.02$ respectively, from Table 1 of T+14, vs. $0.28\pm0.06$ from out Table \ref{tab:tab2}), while a small offset in mass is also visible due to difference in mass inferences. A more significant tilt of the slopes of both the N+11 and SPIDER samples is evident. This might have some different origins. For N+11, they do not discuss the use of an aperture correction, hence part of the tilt might come from the aperture correction. For SPIDER, their analysis is more focused on the massive sample, hence their slope is more sensitive to the region of the relationship that shows a knee, at $\log M_*/M_\odot\gsim 10.6$.

For LTGs there are little reference in literature to compare our results to: for instance, we have found that our best fit parameters in Table \ref{tab:tab2} lie in between best fit Mass--$\sigma$ relations by \citet[][e.g. their Table 3]{2019MNRAS.489.3797M} from 34 nearby spirals from the Calar Alto Legacy Integral Field Area (CALIFA) survey.  

Curvatures and bendings in galaxy scaling relations are emerging
at some characteristic mass scales. Virial mass (e.g.
\citealt{Moster+10}), size \citep{HB09_curv}, central DM fraction
and total mass density slope
\citep{Cappellari+13_ATLAS3D_XX,Tortora+19_LTGs_DM_and_slopes},
optical colour, metallicity and stellar M/L gradient
\citep{Spolaor+10,Tortora+10CG,Tortora+11MtoLgrad} present
bendings or upturn/downturn at $\mst \sim 3 \times 10^{10}\, \rm
M_{\rm \odot}$ (Chabrier IMF).  Such correlations are shaped by
different kinds of physical processes which can be regulated by
this characteristic mass
(\citealt{Tortora+19_LTGs_DM_and_slopes}).

\begin{figure}
\hspace{-0.5cm}
        \includegraphics[width=1.02\columnwidth]{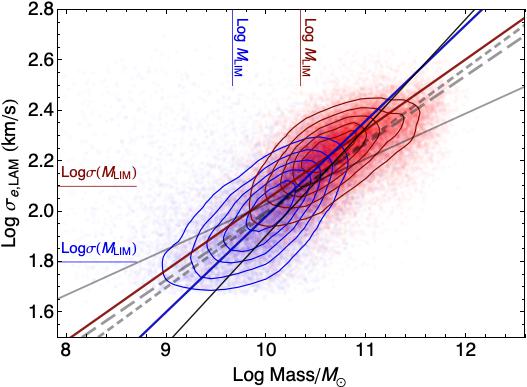}
    \caption{Mass--$\sigma$ relation from LAMOST galaxies from Fig. \ref{Fig:mass_sigma1}: ETGs (red dots) and LTGs (blue dots). Galaxies with different $n-$index show a different slope as demonstrated by the red and blue solid lines. The slope of the ETG sample is nicely consistent with local samples from \citet{Tortora+14_DMevol} (short and long dashed gray lines) but slightly tilted with respect to the SPIDER/SDSS sample from \citet{SPIDER-VI} (solid gray line) and \citet{2011A&A...534A..61N} (solid black line).}
    \label{fig:mass_sigma2}
\end{figure}

Our Mass-$\sigma$
%, and in particular in the overall Mass-$\sigma$
relation also shows a bending signature, which is more evident in
Fig. \ref{fig:mass_sigma2}, where we combine all results from this
work and previous literature relations. Here we see that the two
samples of galaxies, ETGs and LTGs, albeit separated
simplistically through their $n-$index, show two main features: 1)
they seem to lie on a similar relation at masses lower than  $\log
M_*/M_\odot\gsim10.6$ (Kroupa IMF), while ETGs sit on a shallower
relation at masses higher than  $\log M_*/M_\odot\gsim10.6$, which
determines the shallower overall slope found in Table
\ref{tab:tab2} and drawn by the best--fit curve. If one would
select a more massive sample, an even shallower slope might be
found (see e.g. \citealt{SPIDER-VI}). This will be investigated
in mode details in forthcoming analyses.

Indeed, a full discussion of the Mass-$\sigma$ relation is beyond
the purpose of this paper, which is partially based on publicly
available data (e.g. masses) which we cannot fully control.
%Hence, we focus here on the general check that the results implied from the LAMOST VD measurement are consistent with previous findings.

%Also, they do not discuss the use of an aperture correction, hence we expect that this might introduce a tilt in their relation.
%For instance, in Finally they derived their relation for a local sample of galaxies ($0.01<z<0.08$), and also if the LAMOST galaxies are distributed on a wider range of redshifts (i.e. $z\sim0.3$, see e.g. Fig.\ref{fig:LAMOST_z}), the different redshift window adopted is also a source of systematics between the two samples.
Here we just remark that, despite all inhomogeneities with other
datasets, the fact that the best fit relations are very similar
and statistically consistent (in the range of mass above the
limiting mass of the LAMOST sample) with external datasets
demonstrates that LAMOST VD estimates are robust for further
science exploitation.

\section{Conclusions}
\label{sec:conclusions}
We have produced the first catalog of velocity dispersion measurements for galaxies belonging to the LEGAS sample observed with LAMOST. The catalog is based on the spectra delivered with the LAMOST-DR7. This includes 193\,361 spectra classified as galaxy-type which are distributed over $\sim 11\,500$ deg$^2$ of the sky and overlap with other imaging (SDSS, HSC) and spectroscopic (BOSS) surveys. After a quality selection, largely based on a conservative cut in SNR in $g-$ and $i-$ band. We have adopted a wrap-up procedure to perform the spectral fitting using \textsc{pPXF}, and derive velocity dispersion measurements. After applying further quality criteria, like signal-to-the-residual ratio (SRR)>5, the $\chi^2_{\rm FIT}$ < 1.1 and a velocity dispersion range of 50 km/s $\leq\sigma_{\rm meas}\leq$ 450 km/s, we have collected central velocity dispersion measurements of $\sim86\,000$ galaxies,
%included in the data releases until 7th,
which we make publicly available. About 50\,500 of these galaxies
have a match with SDSS and $\sim$2600 with BOSS, and have
structural parameters (e.g. S\'ersic Index, effective radius,
Petrosian magnitudes) and stellar masses available from literature
(e.g. NYU-VAGC, SDSS). For the other galaxies similar quantities
need to be derived from suitable ancillary datasets.

We have estimated statistical errors as a function of the SNR
using both duplicated spectra of individual galaxies from LAMOST
(Fig. \ref{fig:rel_err}) and compared the relative scatter of
LAMOST estimates against a common sample of SDSS and BOSS galaxies
(Fig. \ref{fig:sigma_lamost_sdss} and Fig.
\ref{fig:sigma_lamost_boss}) and they all consistently show that
internal statistical errors are always within 20\% at SNR$>20$ and
down to $\lsim$15\% for SNR$>50$. In the same figures, we can also
see that after having corrected the VD estimate for fiber
apertures and computed the velocity dispersion at the effective
radius, we have found almost no systematics between the LAMOST
estimates and SDSS and BOSS, as a function of both the LAMOST
velocity dispersion values and the SNR.

Finally as a scientific validation of the new VD catalog, in combination with other ancillary data, we have derived the Mass$-\sigma$ relation of the LAMOST dataset after having separated early-type galaxies (ETGs) from late-type galaxies (LTGs) solely on the basis of their S\'ersic-index, and compared with other literature studies. We have found a substantial agreement of the slope of the relation of the ETGs with other local samples (see Fig. \ref{fig:mass_sigma2}), largely based on SDSS data, from \citet{2011A&A...534A..61N} and \citet{Tortora+14_DMevol}, while LTGs show a steeper slope (also in Fig. \ref{fig:mass_sigma2}, but see also Tab \ref{tab:tab2}), although significant only at about 2$\sigma$ level.

We plan to develop the machinery presented here with a more sophisticated analysis, implementing 1) a resampling the LAMOST spectra and 2) a bootstrap of \textsc{pPXF} set-up in order to associate more realistic errors to the VD estimates and to minimize the source of systematics.

With this first analysis we have demonstrated that LAMOST spectra are suitable for galaxy kinematics. This represents a value added product for the LAMOST project but has also a legacy value because it has a large overlap with SDSS/BOSS in the NGC, and provides first VD estimates for galaxies in a large area in the SGC, which will be soon targeted by optical/NIR space facilities like EUCLID and the Chinese Space Station Telescope.

\section*{Acknowledgements}
Guoshoujing Telescope (the Large Sky Area Multi-Object Fiber Spectroscopic Telescope, LAMOST) is a National Major Scientific Project which is built by the Chinese Academy of Sciences, funded by theNational Development and Reform Commission, and operated and managed by the NationalAstronomical Observatories, Chinese Academy of Sciences.\\
NRN acknowledges financial support from the “One hundred top talent program of Sun Yat-sen University” grant N. 71000-18841229, and from the European Union Horizon 2020 research and innovation programme under the Marie Skodowska-Curie grant agreement n. 721463 to the SUNDIAL ITN network. GD acknowledges support from CONICYT project Basal AFB-170002. CT acknowledges funding from the INAF PRIN-SKA 2017 program 1.05.01.88.04. GZ and ALL acknowledge the support by the National Natural Science Foundation of China under grant Nos. 11988101, U1931209 and the National Key R\&D Program of China No. 2019YFA0405502.

\subsubsection*{Data Availability}
The velocity dispersion catalog produced within this article is available at the following URLs:  http://dr7.lamost.org/v1.1/doc/vac (LAMOST website) and  http://paperdata.china-vo.org/SYSU/LAMOST\_DR7\_VDcat/lamost\_DR7\_VDcat.zip (China-VO direct download).

%%%%%%%%%%%%%%%%%%%%%%%%%%%%%%%%%%%%%%%%%%%%%%%%%%



\bibliographystyle{mnras}
%\bibliography{LAMOST_bibliography} % if your bibtex file is called example.bib

\input{LAMOST_VD_ArXiV.bbl}

%\begin{thebibliography}{99}
%\end{thebibliography}

%%%%%%%%%%%%%%%%%%%%%%%%%%%%%%%%%%%%%%%%%%%%%%%%%%

%%%%%%%%%%%%%%%%% APPENDICES %%%%%%%%%%%%%%%%%%%%%

\appendix

\section{Preliminary tests on \textsc{pPXF} bootstrapping and template mismatch}\label{sec:app1}
{In the \S\ref{templates_rebin}, we have illustrated the \textsc{pPXF} set-up adopted for this first VD catalog from LAMOST spectra and anticipated what are the pipeline improvements that we plan to implement for future analyses. In particular we plan to obtain average estimates and realistic statistical errors by bootstrapping over different \textsc{pPXF} set-up and resampling the LAMOST spectra using the spectral noise. As a first quick test on the outcome of this analysis, we have experimented a preliminary procedure over a limited sample.

The procedure randomly creates 256 realizations of each observed spectrum by adding Gaussian noise to each pixel and by randomly shifting (to bluer or redder wavelengths) the selected restframe wavelength window of the spectrum by a maximum of 10 pixels. It also varies a series of fitting parameters like %the subset of stellar templates to use for the fit among the full library of templates,
the LOSVD moments, the degree of additive polynomials, and the initial guesses of radial velocity and velocity dispersion, (see D'Ago et al., in preparation).

In Fig. \ref{fig:appendix1}, we demonstrate the results for $\sim200$ galaxies, where the single \textsc{pPXF} measurements are plotted against the bootstrap average value (errorbars are the standard deviation over the 256 realizations). As it is seen the estimates nicely sit on the one-to-one correlation (solid line) with average relative scatter of $\sim$0.13, e.g. consistent with the one found in the repeated spectra experiment in Figs. \ref{fig:rel_err} and \ref{fig:best_duplic} (see \S\ref{sec:repeated}).
The reason of such a very good accuracy of the single measurement, despite some simplistic assumptions, e.g. about initial conditions and the number of templates (i.e. the use of 40 MILES SSP models from \citealt{Vazdekis2010}, see also below) resides on the quite high SNR and good resolution of the LAMOST spectra. The full bootstrap procedure, though, will return more realistic errors associated to the VD values. Also, it will be necessary to assess internal statistical errors and systematics in case we will adopt looser constraint on the SNR spectra quality overall, to increase the completeness of the VD catalog in future releases.

\begin{figure}
    \centering
        \includegraphics[width=0.9\columnwidth]{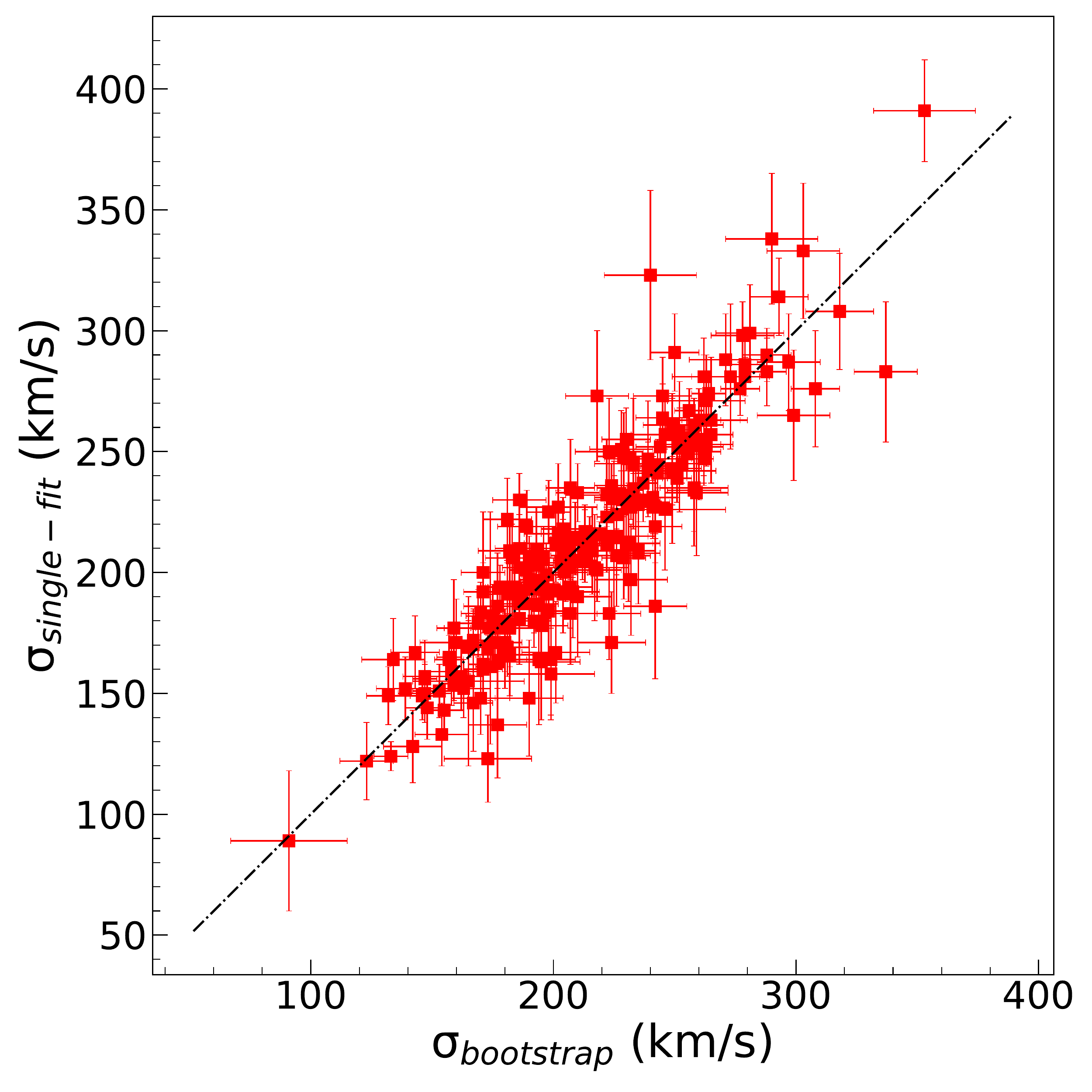}
    \caption{Preliminary bootstrap test over a sample of $\sim200$ LAMOST galaxies. Comparison between the single run \textsc{pPXF} VD estimates and the ones obtained over the bootstrap procedure where 256 are averaged after LAMOST spectra are resampled, the \textsc{pPXF} parameters are randomly changed and the stellar templates are also picked across the MILES catalog. See text for more details. The solid line represents the one-to-one relation.}
    \label{fig:appendix1}
\end{figure}

Finally, a note about the template choice. Stellar template mismatch is one limiting factor for velocity dispersion accuracy. % The randomization of templates in the bootstrap procedure can account for internal statistical errors, but not for systematics related to the template models and/or assumptions.
In the current analyses we have used a selection of 40 MILES SSP models, covering a wide range of metallicities (0.02$\leq Z/Z_\odot\leq$1.58) and ages (between 1 Gyr and 13 Gyr). In principle we could have used single star templates (e.g. among the 268 empirical stars from MILES library, uniformly sampling effective temperature, metallicity and surface gravity of the full catalogue of templates), but at the cost of a much longer computing time.
However, we have checked in previous analyses (e.g. \citealt{Scognamiglio+20_UCMGs}) that this does not affect considerably the VD results. However, we will address this effect more in details in future analyses.}

\section{Redshift consistency}\label{sec:app2}
{Source redshifts are the basic entries of the LAMOST data releases, for all spectra and classes of objects. This paper aims at providing the first catalog of velocity dispersion estimates as an additional kinematical parameter to the standard LAMOST catalog entries. Of course, when performing the template fitting to the spectra, the assumed velocity distribution (which we have set to be Gaussian in this first VD release) incorporates the mean redshift as a fitting parameter. Hence, we provide, as an output of the VD catalog, the difference between the LAMOST pipeline released redshift and the estimated \textsc{pPXF} one, $\Delta z$ (see \S\ref{sec:catalog_content}) for all galaxies with SNR$_i>10$ or SNR$_g>10$. For all other galaxies (and other source types) with SNR<10, the redshifts are still retrievable by the LAMOST-DR7 general release.

\begin{figure}
\hspace{-0.5cm}
        \includegraphics[width=1.05\columnwidth]{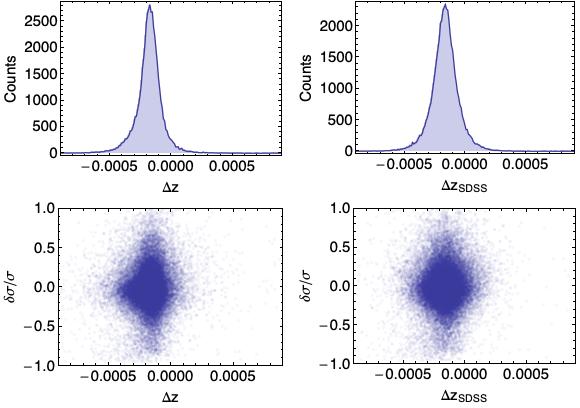}
    \caption{Redshift consistency between the \textsc{pPXF} estimates and the ones of the original LAMOST catalog ($\Delta z$), and the redshifts from SDSS and BOSS systems matching with the LAMOST VD catalog ($\Delta z_{\rm SDSS}$). {\it Top row.} The ($\Delta z$) distribution (left) and the $\Delta z_{\rm SDSS}$ (right) showing an offset with respect to zero. {\it Bottom row.} The relative scatter of the LAMOST VD estimates (as defined in \S\ref{sec:tests}) versus $\Delta z$ (left) and $\Delta z_{\rm SDSS}$ (right) showing no evident correlations in both cases.}
    \label{fig:appendix2}
\end{figure}

For sake of completeness, in this appendix, we briefly quantify the $\Delta z$, introduced in \S\ref{sec:catalog_content}, although any detailed discussion of discrepancies among pipelines is beyond the purpose of this paper, unless they affect the VD estimates. The distribution of the $\Delta z$ is reported in Fig. \ref{fig:appendix2} (top-left). Here we see that $\Delta z$ are oddly distributed with respect to $\Delta z=0$, with a median value of the $\langle \Delta z \rangle=-0.00018\pm0.00009$, i.e. $\sim 2\sigma$ larger than zero, meaning that there is a significant offset. The origin of this systematic difference remains unclarified, and it is also present in the $\Delta z_{\rm SDSS}$ with SDSS/BOSS (defined as $\Delta z_{\rm SDSS}=z_{\rm SDSS} - Z_{\rm ppxf}$ and also shown in Fig. \ref{fig:appendix2}), which is $-0.00017\pm0.00010$. This means that this systematic effect has been introduced by the \textsc{pPXF} procedure. However, this is not worrisome as long as we can demonstrate that it does not introduce any systematic on the VD estimates.

In the same Fig. \ref{fig:appendix2} we plot the VD scatter of the LAMOST/\textsc{pPXF} estimates vs. SDSS/BOSS, as a function of the $\Delta z_{\rm LAM}$ and $\Delta z_{\rm SDSS}$. We do not measure any correlation in both cases, which means that the overall VD estimates reported in the catalog are unaffected by this offset, which we plan to solve in the next releases.}

%%%%%%%%%%%%%%%%%%%%%%%%%%%%%%%%%%%%%%%%%%%%%%%%%%

% Don't change these lines
\bsp    % typesetting comment
\label{lastpage}
\end{document}